\documentclass[journal,twoside,nofonttune]{IEEEtran}

\usepackage{amsmath}
\usepackage{amsfonts}
\usepackage{amssymb}
\usepackage{mathtools}

\usepackage{float}
\usepackage{color}
\usepackage{graphicx}
\usepackage[protrusion = true]{microtype}
\usepackage[subtle, mathspacing = normal]{savetrees}
\usepackage{bm}

\usepackage[font=small]{caption}
\usepackage{subcaption}

\usepackage{balance}

\usepackage{algorithm}
\usepackage{algpseudocode}

\newcommand{\Input}{\State\textbf{Input: }}
\newcommand{\Output}{\State\textbf{Output: }}

\usepackage[utf8x]{inputenc} 
\usepackage[english]{babel}

\DeclareMathOperator*{\argmax}{argmax}

\def\A{{\mathbf A}}

\def\R{{\mathbf R}}
\def\P{{\mathbf P}}

\def\Q{{\mathbf Q}}

\def\G{{\mathbf G}}
\def\H{{\mathbf H}}

\def\M{{\mathbf M}}

\def\x{{\mathbf x}}

\def\y{{\mathbf y}}

\def\Id{\mathbf{Id}}
\newcommand{\btheta}{\bm{\theta}}
\newcommand{\Real}{\mathbb{R}}

\newcommand{\stcomp}{\mathsf{c}}

\usepackage{multirow}

\makeatletter
\def\blfootnote{\xdef\@thefnmark{}\@footnotetext}
\makeatother

\title{Sparse Bayesian Estimation of Parameters in Linear-Gaussian State-Space Models}
\author{Benjamin~Cox,~\IEEEmembership{Student Member,~IEEE} and
	V\'ictor~Elvira,~\IEEEmembership{Senior Member,~IEEE}
}

\begin{document}
\maketitle
\blfootnote{B.C. acknowledges support from the \emph{Natural Environment Research Council} of the UK through a SENSE CDT studentship (NE/T00939X/1). The work of V. E. is supported by the \emph{Agence Nationale de la Recherche} of France under PISCES (ANR-17-CE40-0031-01), the Leverhulme Research Fellowship (RF-2021-593), and by ARL/ARO under grants W911NF-20-1-0126 and W911NF-22-1-0235. }
\begin{abstract}

State-space models (SSMs) are a powerful statistical tool for modelling time-varying systems via a latent state. 
In these models, the latent state is never directly observed. 
{Instead,} a sequence of {data points} related to the state are {obtained}. 
{The linear-Gaussian state-space model is widely used, since it allows} for exact inference when all model parameters are known, {however this is rarely the case}. 
{The estimation of these parameters is a very challenging but essential task to perform inference and prediction.}
{In the linear-Gaussian model}, the state dynamics are described via a state transition matrix. 
This model parameter is known to behard to estimate, since it encodes the relationships between the state elements, which are never observed. 
{In many applications, this transition matrix is sparse since not all state components directly affect all other state components.} 
However, most parameter estimation methods do not exploit this feature. 
{In this work we propose SpaRJ, a fully probabilistic Bayesian approach that obtains sparse samples from the posterior distribution of the transition matrix.}  
Our method explores sparsity by traversing a set of models that exhibit differing sparsity patterns in the transition matrix.
Moreover, we also design new effective rules to explore transition matrices within the same level of sparsity.
This novel methodology has strong theoretical guarantees, and unveils the latent structure of the data generating process, thereby enhancing interpretability. The performance of SpaRJ is showcased in example with dimension 144 in the parameter space, and in a numerical example with real data.
\end{abstract}


\section{Introduction}

\IEEEPARstart{S}{tate}-space models (SSMs) are a flexible statistical framework for the probabilistic description of time-variable systems via coupled series of hidden states and associated observations. 
{These models are used to decode motor neuron kinematics from hand movements \cite{aghagolzadeh2014latent}, perform epidemiological forecasting for policy makers \cite{scienceImperial, WoodPLOS}, and to determine the trajectory of lunar spacecraft from noisy telemetry data \cite{grewal2010applications}, among many other applications.
In general, SSMs are used in many important problems within, but not limited to, signal processing, statistics, and econometrics \cite{hamilton1986standard}. }
{
In some cases, the true SSM is perfectly {known, and the interest is in inferring the sequence of underlying hidden states.}
In the Bayesian paradigm, estimating the sequence of hidden states is achieved by obtaining a sequence of posterior distributions of the hidden state, known as filtering distributions \cite{sarkka2013bayesian}.
The linear-Gaussian SSM (LGSSM) is the case where the state and observation models are linear with Gaussian noises. 
In this case, the sequence of exact filtering distributions is obtained via the Kalman filtering equations \cite{kalman1960new}.
In the case of non-linear dynamics, the filtering distributions must be approximated, for instance via the extended Kalman filter \cite{einicke1999robust} or the unscented Kalman filter \cite{wan2000unscented}.}
{In even more generic models, such as those with non-Gaussian noise, particle filters (PFs) are often used,} {which approximate the state posterior distributions via Monte Carlo samples \cite{Gordon93,djuric2003particle,doucet2009tutorial,elvira2019elucidating,branchini2021optimized}.}
All of these filtering methods assume that the model parameters are known.
{
However, model parameters are often unknown, and must therefore be estimated.}
{Parameter estimation is a much more difficult task than filtering, and is also more computationally expensive, since most parameter estimation algorithms require a large number of evaluations of the filtering equations to yield acceptable parameter estimates.}
{There exist several generic methods to do this, with techniques based on Markov chain Monte Carlo (MCMC) \cite{andrieu2010particle} and expectation maximisation (EM) \cite{sarkka2013bayesian} arguably being the most commonly used parameter estimation techniques for state-space models.}

{When estimating model parameters, it is crucial that the estimates reflect the structure of the underlying system.
For instance, in real-world systems, the underlying dynamics are often composed of simple units, with each unit interacting with only a subset of the overall system, but when observed together these units exhibit complex behaviour \cite{watts1998collective}.}
{This structure can be recovered by promoting sparsity in the parameter estimates.
In addition to better representing the underlying system, sparse parameter estimates have several other advantages.}
By promoting sparsity uninformative terms are removed from the inference, thereby reducing the dimension of the parameter space, improving model interpretability.
Furthermore, parameter sparsity allows us to infer the connectivity of the state space \cite{chouzenoux2020graphem}, which is useful in several applications, such as biology \cite{pirayre2018brane, luengo2019hierarchical}, social networks \cite{ravazzi2017learning}, and neuroscience \cite{richiardi2013machine}.
In state-space models the sparsity structure can be represented as a directed graph, with the nodes signifying the state variables, and edges indicating signifying between variables.
{In the LGSSM specifically, this graph can be represented by an adjacency matrix with identical sparsity to the transition matrix.} 
{This interpretation of a sparse transition matrix as a weighted directed graph was recently proposed in the GraphEM algorithm \cite{chouzenoux2020graphem,elvira2022graphical,chouzenoux2023graphit}, in which a sparse point-wise maximum-a-posteriori estimator for the transition matrix of the LGSSM is obtained via an EM algorithm.}
{However, this point estimator does not quantify uncertainty, therefore disallowing a probabilistic evaluation of sparsity. }
{The capability to quantify and propagate the uncertainty of an estimate is highly desired in modern applications, as it allows for more informed decision-making processes, as well as providing a better understanding of the underlying model dynamics.}

{
In this work, we propose the \emph{sparse reversible jump} (SpaRJ) algorithm, a fully Bayesian method to estimate the state transition matrix in LGSSMs.
This matrix is probabilistically approximated by a stochastic measure constructed from samples obtained from the posterior of this model parameter.}
{
The method (a) promotes sparsity in the transition matrix, (b) quantifies the uncertainty, including sparsity uncertainty in each element of the transition matrix, and (c) provides a probabilistic interpretation of (order one) Granger causality between the hidden state dimensions, which is interpreted as a probabilistic network of how the information flows between consecutive time steps.
}

{
SpaRJ exploits desirable structure properties within the SSM, which presents computational advantages w.r.t. to other well established MCMC methods such as particle MCMC \cite{andrieu2010particle}, i.e., a decrease in computational cost for a given performance or an improvement in performance for a given computational cost. 
}

{
Our method is built on a novel interpretation of sparsity in the transition matrix as a model constraint. SpaRJ belongs to the family of reversible jump Markov chain Monte Carlo (RJMCMC) \cite{green1995reversible,cappe2003reversible}, a framework for the simultaneous sampling of both model and parameter spaces.
We note that RJMCMC methods are not a single algorithm, but a wide family of methods (as it is the case of MCMC methods). Thus, specific algorithms are required to make significant design choices so the RJMCMC approach can be applied in different scenarios \cite{green1995reversible, cappe2003reversible, robert2013monte, richardson1997bayesian}.
In the case of SpaRJ, we design both specific transition kernels and parameter rejuvenation schemes, so the algorithm can efficiently explore both the parameter space, and the sparsity of the parameter in a hierarchical fashion.
As RJMCMC is itself a modified Metropolis-Hastings method, the solid theoretical guarantees of both precursors are inherited by our proposed algorithm, such as the asymptotic correctness of distribution of both model and parameter \cite{green1995reversible}.
}
{Our method outperforms the current state-of-the-art methods in two numerical experiments. We test SpaRJ in a synthetic example with dimension up to 144 in the parameter space. In this example, a total of $2^{144}$ models are to be explored (i.e., the number of different sparsity levels). Then, we run a numerical example with real data of time series measuring daily temperature. The novel probabilistic graphical interpretation allows recovery of a probabilistic (Granger) causal graph, showcasing the large impact that this novel approach can have in relevant applications of science and engineering.
The model transition kernels used by SpaRJ are designed to allow the exploitation of sparse structures that are common in many applications, which reduces the computational complexity of the resulting (sparse) models once the transition matrix has been estimated (see for instance \cite{watts1998collective}).
SpaRJ retains strong theoretical guarantees, inherited from the underlying Metropolis-Hastings method, thanks to careful design of the transitions kernels, e.g., keeping the convergence properties of the algorithm.
Extending our methodology to parameters other than the transition matrix is readily possible. In particular, we make explicit both a model and parameter proposal for extension to the state covariance parameter $\Q$.}

\noindent\textbf{{Contributions.}} 
{The main contributions of this paper\footnote{{A limited version of this work was presented by the authors in the conference paper \cite{cox2022parameter}, which contains a simpler version of the method with no theoretical discussion, methodological insights, or exhaustive numerical validation.}} are summarised as follows:}
\begin{itemize}
    \item The proposed SpaRJ algorithm is the first method to estimate probabilistically the state transition matrix in LGSSMs {(i.e, treating $\A$ as a random variable rather than a fixed unknown)} under sparsity constraints. {This is achieved by taking $\A$ to be a random variable, and sampling the posterior distribution $p(\A|\y_{1:T})$ under a unique interpretation of sparsity as a model. This new capability allows for powerful inference to be performed with enhanced interpretability in this relevant model}, e.g., the construction of a probabilistic Granger causal network mapping the state space, which was not possible before. 
    \item The proposed method {is the first method to} quantify the uncertainty associated with the occurrence of sparsity {in SSMs}, e.g., in the probability of sparsity occurring in a given element of the transition matrix. This capability is unique among parameter estimation techniques in this field.
    \item {Our method proposes an interpretation of sparsity as a model, allowing the use of RJMCMC for sparsity detection in state-space modelling. 
    This is the first RJMCMC method to have been applied to sparsity recovery in state-space models, probably because RJMCMC methods require careful design of several parts of the algorithm, especially for high dimensional parameter spaces as is the case for the matrix valued parameters of the LGSSM.}
\end{itemize}

\noindent\textbf{Structure.} In Section~\ref{sec:background} we present the components of the problem and present some of the underlying algorithms, as well as the notation we will use.
Section~\ref{sec:sparj} presents the method, with further elucidation in Section~\ref{sec:sparjdisc}.
We present several challenging numerical experiments in Section~\ref{sec:numerical}, showcasing the performance of our method, and comparing to a recent method with similar goals.
We provide some concluding remarks in Section~\ref{sec:conclusion}.

\section{Background}
\label{sec:background}
\subsection{State-space models}
Let us consider the additive linear-Gaussian state-space model (LGSSM), given by
\begin{equation}
	\begin{aligned}
		\label{eq:lgssm}
		\mathbf{x}_t &= \mathbf{A} \mathbf{x}_{t-1} + \mathbf{q}_t,\\
		\mathbf{y}_t &= \mathbf{H} \mathbf{x}_t + \mathbf{r}_t,
	\end{aligned}
\end{equation}
for $t = 1, \dots, T$, where $\x_t \in \Real^{d_x}$ is the hidden state with associated observation $\mathbf{y}_t \in \Real^{d_y}$ at time $t$, $\mathbf{A} \in \Real^{d_x \times d_x}$ is the state transition matrix, $\mathbf{H} \in \Real^{d_y \times d_x}$ is the observation matrix, $\mathbf{q}_t \sim \mathcal{N}(\mathbf{0}, \mathbf{Q})$ is the state noise, and $\mathbf{r}_t \sim \mathcal{N}(\mathbf{0}, \mathbf{R})$ is the observation noise.
The state prior is $\mathbf{x}_0 \sim \mathcal{N}(\bar{\mathbf{x}}_0, \mathbf{P}_0)$, with $\bar{\mathbf{x}}_0$ and $\mathbf{P}_0$ known. {We assume that the model parameters remain fixed.}

{A common task in state-space modelling is the estimation of the series of $p(\x_t|\y_{1:t})$ for $t \in \{1, \dots, T\}$, also known as the filtering distributions.}
{In the case of the LGSSM, these distributions are obtained exactly via the Kalman filter equations \cite{sarkka2013bayesian, kalman1960new}.}
The linear-Gaussian assumption is not overly restrictive, as many systems can be approximated via linearisation, and for continuous problems Gaussian noises are very common.

Note that, the posterior distribution of any given parameter can be factorised as 
\begin{equation}
\label{eq:parampost}
    p(\btheta|\mathbf{y}_{1:T}) \propto p(\mathbf{y}_{1:T}|\btheta) \ p(\btheta),
\end{equation}
where $\btheta$ is the parameter of interest, $p(\btheta)$ is the prior ascribed to the parameter, and $p(\mathbf{y}_{1:T}|\btheta)$ is extracted from the Kalman filter via the recursion
\begin{equation}
\label{eq:ylikdecom}
     p(\mathbf{y}_{1:T}|\btheta) = \prod_{t=1}^{T} p(\mathbf{y}_t|\mathbf{y}_{1:t-1}, \btheta),
\end{equation}
where $p(\mathbf{y}_1|\mathbf{y}_{1:0}, \btheta) := p(\mathbf{y}_1|\btheta)$ \cite{sarkka2013bayesian}.
This factorisation {gives} the target distribution for estimating parameters in an LGSSM.
{In this work we focus on probabilistically estimating $\A$, and therefore we are interested in the posterior $p(\A|\mathbf{y}_{1:T}) \propto p(\mathbf{y}_{1:T}|\A) p(\A)$.}

In LGSSMs, $\H$ and $\R$ are frequently assumed to be known as parameters of the observation instrument, but $\Q$ and $\A$ are often unknown.
For the purposes of this work, we assume that all parameters except $\A$ are known, or are suitably estimated, although our method can be extended to all parameters {of the linear-Gaussian state-space model}, as discussed in Section~\ref{sec:extend}.

\subsection{Parameter estimation in SSMs}
The estimation of the parameters of a state-space model is, in general, a difficult and computationally intensive task \cite{sarkka2013bayesian, andrieu2010particle}.
This difficulty follows from the state dynamics not being directly observed, and stochasticity in the observations.

In this work we focus on Bayesian techniques, as our method is Bayesian, with this focus therefore allowing easier comparison.
Frequentist methods for parameter estimation in SSMs are common however, with some relevant references being \cite{kantas2015particle, doucet2003parameter, campillo2009convolution}. 

There are two main approaches to estimating and summarising parameters in state-space models, which we can broadly classify as point estimation methods and probabilistic methods.

\noindent\textbf{Point estimation methods.} The goal of a point estimation method is to find a single parameter value that is, in some way, the value that best summarises the parameter given the data.
An archetypal point estimate is the maximum likelihood estimator (MLE) \cite{eliason1993maximum}. 
In a state-space model, when estimating a parameter denoted $\btheta$, the MLE, denoted $\hat{\btheta}_{\text{MLE}}$, is given by
\begin{equation*}
	\hat{\btheta}_{\text{MLE}} = \argmax_{\btheta} \ p({\y_{1:T}|\btheta}).
\end{equation*} 
{The MLE is fundamentally a frequentist estimator, and hence no prior distribution is used. The Bayesian equivalent to the maximum likelihood estimator is the maximum a posteriori estimator, denoted $\hat{\btheta}_{\text{MAP}}$, and given by}
\begin{equation*}
	\hat{\btheta}_{\text{MAP}} = \argmax_{\btheta} \ p({\btheta|\y_{1:T}}) = \argmax_{\btheta}\left( p({\y_{1:T}|\btheta})\,p(\btheta)\right),
\end{equation*} 
from which we see that the MLE is the MAP if $p(\btheta) \propto 1$.

A common method for point estimation in LGSSMs {and in general} is the expectation-maximisation (EM) algorithm \cite{dempster1977maximum}, as explicit formulae exist for the conditional MLE of all parameters in the case of the LGSSM, and thus the model parameters can be estimated iteratively.
The EM algorithm allows for all model parameters to be estimated simultaneously, but converges much more slowly as the number parameters to estimate increases \cite{sarkka2013bayesian}.
Furthermore, this method does not allow for quantification of the uncertainty in the resultant estimates.

\noindent\textbf{Probabilistic methods.} 
Distributional methods estimate the target probability density function (pdf) of the parameter given the data, often through the generation of Monte Carlo samples.
In the case of state-space models, for a parameter $\btheta$ the target distribution is $p(\btheta|\y_{1:T})$, and the set of Monte Carlo samples is $\{\btheta_i\}_{i=1}^n$, with $\btheta_i \sim p(\btheta|\y_{1:T}).$
A common class of methods used to obtain these samples is Markov chain Monte Carlo (MCMC), with methods such as particle MCMC \cite{andrieu2010particle} seeing wide use in SSMs.
MCMC is a class of sampling methods that construct, and subsequently sample from, a Markov chain that has the target as its equilibrium distribution \cite{robert2013monte}.
The elements of this chain are then taken to be Monte Carlo samples from the target distribution, although typically a number of the initial samples are discarded to ensure that the samples used are from after the chain has converged \cite{robert2013monte}. 
	
Note that there exist probabilistic methods, such as Laplace approximation \cite{rue2009approximate} and variational inference \cite{blei2017variational}, that provide analytical approximations and do not directly give samples.
These methods are seldom used in state-space modelling. 
Distributional methods give more flexibility than point estimates, as they capture distributional behaviour and inherently quantify uncertainty, as well as allowing point estimates to be estimated from the samples, such as the aforementioned MAP estimator being the maximising argument for the posterior likelihood.

\subsection{Sparse modelling}
When fitting and designing statistical models, the presence of sparsity in parameters is often desirable, as it reduces the number of relevant variables thus easing interpretation and simplifying inference.
Furthermore, real systems are often made up of several interacting dense blocks that, when taken as a whole, exhibit complex dynamics \cite{watts1998collective}.
Sparse estimation methods allow for this structure to be recovered, resulting in estimates that can reflect the structure of the underlying system.
{Sparsity is ubiquitous within signal processing, with signal decomposition into a sparse combination of components being very common \cite{scott1998atomic,mohimani2008fast,zayyani2009iterative}, which can be parameterised via model parameters.
Furthermore, within signal processing, there exist a number of existing sparse Bayesian methods, such as \cite{ji2008bayesian, schniter2008fast, zayyani2009bayesian, korki2016iterative}, although these do not operate within the paradigm of state-space modelling. }

There are several approaches to estimate model parameters such that sparsity may be present.
{One approach may be to construct many models with unique combinations of sparse and dense elements, fit all of these models, and then select the best model according to some criteria (see \cite{llorente2020marginal, llorente2022safe, martino2017cooperative} for examples).
This approach is conceptually sound, but computationally expensive for even a small number of parameters $p$, as $2^p$ models must be fitted in order to obtain likelihood estimates, or other goodness-of-fit metrics.
Another approach is to estimate the model parameters under a sparsity inducing penalty, with the classic example of such a penalty being the LASSO \cite{tibshirani1996regression, park2008bayesian}.
This approach, commonly called regularisation, allows for only one model to be fitted rather than many, but increases the computational complexity of fitting the model.
This single regularised estimate is, in most cases, more expensive to compute than fitting a non-regularised estimate as required by the previous approach, but this cost is typically far less than the cumulative cost of all required estimates in the previous approach.
Regularisation is a common way to obtain sparse estimates, and can be extended to Bayesian modelling and estimation in the form of sparsity inducing priors \cite{park2008bayesian, carvalho2010horseshoe}.}

\begin{figure}[ht]
	\centering
	\begin{minipage}{.22\textwidth}
	\centering
	\includegraphics[scale=0.75]{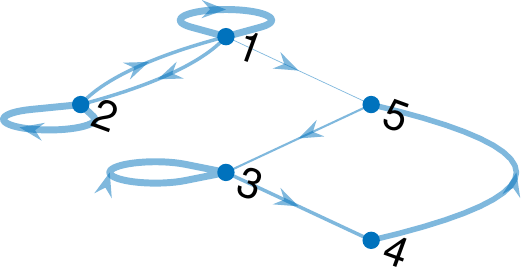}
	\end{minipage}
	\begin{minipage}{.26\textwidth}
	\centering
    \resizebox{1\textwidth}{!}{
	$\A=\begin{pmatrix}
	1 & 0.5 & 0 & 0 & 0\\
	1 & -0.5 & 0 & 0 & 0\\
	0 & 0 & 1 & 0 & -0.3\\
	0 & 0 & 0.1 & 0 & 0\\
	0.1 & 0 & 0 & -0.8 & 0\\
	\end{pmatrix}$}
	\end{minipage}
	
	\caption{Example of a weighted directed graph associated with matrix. The edge thickness corresponds to the magnitude of the weight.}
	\label{fig:adjm_gr}
\end{figure}

In LGSSMs, a sparse estimate of the transition matrix $\A$ can be interpreted as the adjacency matrix of a weighed directed graph $G$, with the nodes being the state elements, and the edges the corresponding elements of $\A$ \cite{chouzenoux2020graphem,elvira2022graphical}. We illustrate this with an example in Fig.~\ref{fig:adjm_gr}.
{We note that there exist a number of graph estimation methods that can be applied to time series data, such as \cite{zheng2018dags, wei2020dags, yu2021dags, bello2022dagma}, although these methods do not utilise the structure of the state-space model. 
Furthermore, these methods often employ an acyclicity constraint, which prevents the results from exhibiting cycles, which are a common feature in real world dynamical systems, for example those resulting from discretised systems of ODEs.}
The graph $G$ thus encodes the linear, between-step relationships of state elements, simplifying model interpretation.
Under this graphical interpretation, $A_{ij}$ being non-zero implies that knowledge of the $j$th element of the state improves the prediction of the $i$th value.
The presence of an edge from node $j$ to node $i$ implies a Granger-causal relationship between the state elements, as knowledge of the past values of the $j$th element at time $t$ improves the prediction of the $i$th element at time $t+1$, which is precisely the definition of a Granger-causal relationship \cite{granger1969investigating}.

\subsection{Reversible jump Markov chain Monte Carlo}

\label{sec:rjmcmc}

Reversible jump Markov chain Monte Carlo (RJMCMC) was proposed as a method for Bayesian model selection \cite{green1995reversible}, and has since seen use in fields such as ecology \cite{pagel2006bayesian}, Gaussian mixture modelling \cite{richardson1997bayesian}, and hidden Markov modelling \cite{cappe2003reversible}.
RJMCMC has even been applied within the realm of signal processing, with some relevant references being \cite{andrieu1999joint, vermaak2004reversible, bunch2016bayesian}.
However, RJMCMC has not been applied to the estimation the sparsity of model parameters within signal processing.

RJMCMC is an extension of the Metropolis-Hastings algorithm that allows for the sampling of a discrete model space, and thus the inclusion of many models within a single sampling chain.
RJMCMC is a hierarchical sampler, with an upper layer sampling the models, and a lower layer sampling the posterior distribution of the parameters within the model. 
This hierarchy allows the use of standard MCMC methods for the lower layer, with the difficulty coming in designing the upper layer \cite{richardson1997bayesian}.

RJMCMC traverses the model space via transition kernels between pairs of models, with the jumps occurring probabilistically. 
This lends the model space an interpretation as a directed graph, with nodes representing the models and edges representing the jumps between models. 
Let $\Theta^{(i)}$ be the parameter space associated with model $M^{(i)}$, and $\btheta^{(i)} \in \Theta^{(i)}$ an associated realisation of the model parameters.
Denote by $\pi_{i,j}$ the probability of jumping from model $M^{(i)}$ to $M^{(j)}$. Note that $\pi_{i,j}$ is zero if and only if $\pi_{j,i}$ is zero. Let $M^{(1)}$ be the current model, and $M^{(2)}$ be a candidate model.
In order to construct a Markov kernel for the transition between models, a symmetry constraint is imposed, i.e., if it is possible to jump from $M^{(1)}$ to $M^{(2)}$, it must also be possible to jump from $M^{(2)}$ to $M^{(1)}$ \cite{robert2013monte}. 
In general however, the dimension of the parameter spaces is not equal, hence it is not possible to construct an invertible mapping between them, violating the required symmetry.
Reversible jump MCMC addresses this by introducing a dimension matching condition \cite{green1995reversible}; the spaces $\Theta^{(1)}$ and $\Theta^{(2)}$ are augmented with simulated draws from selected distributions such that \[(\btheta^{(2)}, u_2) = T_{1,2}(\btheta^{(1)}, u_1), \quad u_1 \sim g_{1,2}(\cdot), \ u_2 \sim g_{2,1}(\cdot),\] where $T_{1,2}$ is a bijection, and $g_{i,j}(\cdot)$ are known distributions. 

The parameter mappings and stochastic draws change the equilibrium distribution of the chain, which means that sampling will not be asymptotically correct.
This counteracted by modifying the acceptance ratio; in the case of jumping from model $M^{(1)}$ to model $M^{(2)}$, the acceptance ratio is given by
\begin{equation}
	\alpha^{(1,2)} = \left|\frac{\partial T_{1,2}(\btheta^{(1)}, u_1)}{\partial(\btheta^{(1)}, u_1)}\right|\frac{g_{2,1}(u_2)}{g_{1,2}(u_1)}\frac{\pi_{2,1}}{\pi_{1,2}}\frac{p_2(\btheta^{(2)})}{p_1(\btheta^{(1)})},
	\label{eq:rjaccrat}
\end{equation}
where $p_i(\btheta^{(i)})$ is the density associated, with model $M^{(i)}$ evaluated at $\btheta^{(i)}$.
The proposal ($M^{(2)}$, $\btheta^{(2)}$) is accepted with probability $\min(\alpha^{(1,2)}, 1)$, and is otherwise rejected.
On rejection, the previous value of the chain is kept, ($M^{(1)}$, $\btheta^{(1)}$).

{In this way, given data $\y$, RJMCMC samples the joint posterior $p(\bm\theta, M|\y)\propto p(M|\x)p(\bm\theta|M, \y)$. 
However, in the case only where only $\bm\theta$ is of interest, following standard Monte Carlo rules, we can discard the samples of $M$ to obtain $p(\bm\theta|\y)$ \cite{robert2013monte}. }

Reversible jump MCMC incorporates many models into a single chain, so it is simple to compare or average models.
However, the parameter mappings and model jump probabilities must be well designed.
Poor selection of these parameters will typically lead to poor mixing in the model space \cite{green1995reversible, robert2013monte}.
Our method explores only a single overall model, which simplifies the mappings. 
We impose a pairwise structure on the model space, simplifying the jumps significantly. 

\subsection{Model definitions and notation}
\label{sec:modnot}
In order to use RJMCMC to explore sparsity in the transition matrix of a linear-Gaussian state-space model, we must first construct a set of candidate sub-models of the LGSSM that exhibit various sparsity levels. To this end we introduce the notation in Table~\ref{tab:notation}.

\begin{table}[H]
	\centering
	\caption{Notation Reference}
	\begin{tabular}{|c|c|}
		\hline
		Notation & Meaning \\
		\hline
		\hline
		$M_n$& Model at iteration $n$\\
		\hline
		$\mathcal{M}_n$& Indices of dense elements in $M_n$\\
		\hline
		$\Theta_n$& Parameter space sampled at iteration $n$\\
		\hline
		$D_n/S_n$& Number of dense/sparse elements at iteration $n$\\
		\hline
	\end{tabular}
	\label{tab:notation}
\end{table}

Denote by $M_n$ the model selected by the algorithm at iteration $n$.
This model is uniquely defined by the associated set of indices of dense elements in $\A$, which we denote $\mathcal{M}_n$.
{Note that there are thus $2^{d_x^2}$ models in our model space, precluding parallel evaluation for even small $d_x$. 
It is therefore not possible to use methods such as Bayes factors or marginal likelihood to compare models, as is standard in Bayesian model selection in state-space models \cite{llorente2020marginal, martino2017cooperative}, as the computational cost is infeasible, due to these requiring all models to be evaluated in order to be compared.}
The number of elements of $\mathcal{M}_n$, denoted by $D_n = |\mathcal{M}_n|$, is the number of dense elements at iteration $n$, and therefore the number of non-zero elements of $\A_n$.
Denote by $S_n = d_x^2 - D_n$ the number of sparse elements at iteration $n$.
If the true value of a parameter is known, then it will be presented without subscript or superscript.
We denote by a superscript $\stcomp$ the complement to a set, and note that $\mathcal{M}^\stcomp_n$ denotes the set of indices of elements sparse in $\A$ at iteration $n$.

Each model has an associated parameter space, which we denote by $\Theta_n$.
As the parameter space of a sparse parameter is $\{0\}$, we therefore have 
\begin{equation}
	\label{eq:zerosparse}
	\Theta_n = \prod_{0<i,j\leq d_x} \btheta_{(i,j),n}, \ \btheta_{(i,j),n} = 
	\begin{cases}
		\Real, & (i,j) \in \mathcal{M}_{n},\\
		\{0\}, & \text{otherwise},\\
	\end{cases}	
\end{equation}
{where $\btheta_{(i,j),n}$ is the support for $(\A_n)_{ij}$.}
In the Bayesian paradigm, we can interpret the sparsity of model $M_n$ as a prior constraint induced by $p(\A|M_n)$, under which the elements indexed by $\mathcal{M}_n^\stcomp$ are always of value $0$, and the elements indexed by $\mathcal{M}_n$ are distributed as $p(\A)$.
This is equivalent to fixing the value of the elements of $\A$ indexed by $\mathcal{M}_n^\stcomp$ to $0$, as in Eq.~\eqref{eq:zerosparse}. 

We denote the $k \times k$ identity matrix by $\Id_k$, and the $k$-vector with all elements equal to $1$ by $\mathbf{1}_k$.
We denote by $\cdot$ any unspecified parameters of a function or distribution.
For example, $x \sim g(\cdot)$ means that the parameters of the distribution $g$ are unspecified, typically as they are irrelevant to the discussion.

{We provide a table of our most used acronyms in Table~\ref{tab:abbrv}.}

\begin{table}[H]
	\centering
	\caption{{List of acronyms}}
	\begin{tabular}{|c|c|}
		\hline
		Abbreviation & Meaning \\
		\hline
		\hline
		MCMC & Markov Chain Monte Carlo\\
		\hline 
		RJMCMC & Reversible Jump MCMC\\
		\hline
		MH & Metropolis Hastings\\
		\hline
		RWMH & Random walk Metropolis Hastings\\
		\hline
		EM & Expectation-Maximisation\\
		\hline
		SSM & State-space model\\
		\hline
		LGSSM & Linear Gaussian state-space model\\
		\hline
	\end{tabular}
	\label{tab:abbrv}
\end{table}

\section{The SpaRJ algorithm}
\label{sec:sparj}

We now present the SpaRJ algorithm, a novel RJMCMC method {to obtain sparse samples from the posterior distribution $p(\A|\y_{1:T})$} of the transition matrix of an LGSSM.
We present our method in Algorithm~\ref{alg:gensparthreestep} for estimating the transition matrix $\A$, although the algorithm can be adapted to estimating any unknown parameter of the LGSSM. 
{Note that our method samples the joint posterior $p(\A, M|\y_{1:t}) \propto p(M|\y_{1:T})p(\A|M, \y_{1:t})p(\A)$ hierarchically, by first sampling $M^\prime$ from $p(M|\y_{1:T})$ and then sampling $\A^\prime$ from $p(\A|M^\prime, \y_{1:t})$ conditional on $M^\prime$.
However, as we are only interested in the posterior of the transition matrix $p(\A|\y_{1:T})$, we marginalise by discarding the samples of $M$ when performing inference \cite{robert2013monte}.}

In order to apply our method, we must provide the values of all known parameters of the LGSSM, (used when evaluating the Kalman filter), and initial values for the unknown parameters $\A$ and $M$.
We initialise the model sampling by setting $M_0$ to the fully dense model. 
The initial value $\A_0$ can be selected in a number of ways, such as randomly or via optimisation \cite{robert2013monte}.
We obtain the initial log-likelihood, $l_0$, by running a Kalman filter with the chosen initial value $\A_0$, giving $l_0 = \log(p(\mathbf{y}_{1:T}|\A_0))$. 
{We define a prior $p(\A)$ on the transition matrix, with some examples given in Section~\ref{sec:priors}.
The prior distribution incorporates our prior knowledge as to the value of the transition matrix $\A$, and can be used to promote sparsity.
However, it is not required in order to recover sparse samples, and can be chosen to be uninformative or diffuse.}

The method iterates $N$ times, each iteration yielding a single sample, outputting $N$ samples, $\{\A_n\}_{n=1}^N$.
While adapting the number of iterations is possible, e.g. by following \cite{dootika2019multivariate}, we present with fixed $N$ so as to provide a simpler algorithm.
Note that, at each iteration we start in model $M_{n-1}$, with transition matrix $\A_{n-1}$, and log-likelihood of $l_{n-1}$.
Each iteration is split into three steps: model proposal (Step 1), parameter proposal (Step 2), and accept-reject (Step 3).
{We note that the model is not fixed, and is sampled at each iteration, allowing for evidence-based recovery of sparsity.}

\noindent\textbf{Step 1: Propose $M^\prime$.} 
At iteration $n$, we retain the previous model $M_{n-1}$ with probability (w.p.) $\pi_0$, and hence setting $M^\prime = M_{n-1}$. 
If a model jump occurs, we set the proposed model $M^\prime$ to be sparser than $M_{n-1}$ w.p. $\pi_{-1}$, and denser otherwise.
To create a model that is sparser, we select a number of dense elements to then make sparse.
To create a denser model, we select a number of sparse elements to make dense.
The number of elements to change, $k$, is drawn from a truncated Poisson distribution (see Appendix~\ref{app:TPoi}) with rate parameter $\lambda_j \in [0,1)$, with this range required in order to bias the jump to models close to the previous model.
This distribution is chosen as it exhibits the required property of an easily scaled support, needed as the maximal jump distance changes with the current model. 
{Note that this distribution does not form a prior over the model space, but instead is used to generate jump kernels, which are then used to explore the model space.
The model space prior $p(M)$ is discussed in Section~\ref{sec:modprior}, and is by default diffuse, i.e. $p(M) = 2^{-d_x^2} \forall M$.}
The truncated Poisson distribution is simple to sample, as it is a special case of the categorical distribution. 
The elements to change are then selected uniformly.
The proposed model $M^\prime$ is always strictly denser than, strictly sparser than, or identical to $M_{n-1}$, following the construction of model jumps in Section~\ref{sec:modadj}. 

\noindent\textbf{Step 2: Propose $\A^\prime$.} 
If the proposed model $M^\prime$ differs from the previous model $M_{n-1}$, then the parameters $\btheta_{n-1}$ are mapped to $\btheta^\prime$ via eq.~\eqref{eq:tfj}, a modified identity mapping.
This mapping is augmented with stochastic draws if the dimension of the parameter space increases, and has elements removed if the dimension decreases.
This mapping has identity Jacobian matrix, and is thus absent from the acceptance ratio.

If the proposed model $M^\prime$ is the same as the previous model $M_{n-1}$, then $A^\prime$ is sampled from the conditional posterior $p(\A|M^\prime)$.
To achieve this, we use a random walk Metropolis-Hastings (RWMH) sampler.
The RWMH sampler requires a single run of the Kalman filter per iteration, which is the most computationally expensive component of the algorithm.
This single run follows from the joint accept-reject decision in Step 3, which allows us to ignore the accept-reject step of a RWMH sampler that jointly assesses all proposals, as in the case of SpaRJ.
We cover the parameter proposal process further in Section~\ref{sec:paramsampling}.
Note that any sampler can be used, even a non-MCMC method, with RWMH chosen for simplicity, computational speed, and to give a baseline statistical performance.

\noindent\textbf{Step 3: Metropolis accept-reject.} 
Once the model and parameter values have been proposed, a Metropolis-Hastings acceptance step is performed. 
We run a Kalman filter with $\A^\prime$ to calculate the log-likelihood of the proposal, $l^\prime = \log(p(\mathbf{y}_{1:T}|\A^\prime))$.

{
Prior knowledge is included via a function of the prior probability densities, denoted $\Lambda$, which encodes both our prior knowledge of the parameter values and of the model (hence the sparsity).
A wide range of prior distributions can be used, with our preference being the Laplace distribution, with the associated $\Lambda$ given in eq.~\eqref{eq:lassolog}, which is known to promote sparsity in Bayesian inference \cite{park2008bayesian}.
Note that the prior is not required to yield sparse samples, but is useful to combat potential over-fitting resulting from the large number of parameters to fit.}
If we denote by  $p(\A)$ our prior on the transition matrix, then $\Lambda(\A_{n-1}, \A^\prime) = \log(p(\A^\prime)) - \log(p(\A_{n-1}))$. 
{In Section \ref{sec:priors}, we provide suggestions as to choosing a prior, and hence $\Lambda(\A_{n-1}, \A^\prime) = \log(p(\A^\prime)) - \log(p(\A_{n-1}))$, }. 

{
When defining the model space in Section~\ref{sec:modnot}, we note that each model is uniquely determined by the sparsity structure it imposes, with this structure being present in all samples of $\A$ generated from this model. 
We can therefore assess the model against our prior knowledge solely based on the sample structure, without a separate prior on the model space.
An example of such a function is the $L_0$-norm, which penalises the number of non-zero elements, a property determined entirely by the model that can be assessed via the samples.
}
{The log-acceptance ratio of the proposed values $\A^\prime$ and $M^\prime$ is given by $\log(a_r) = l^\prime - l_{n-1} + \Lambda^\prime_{n-1} + c$, where $c$ is given in Appendix~\ref{app:corr}.
The model and parameter proposals are jointly accepted with probability $a_r$, and are otherwise rejected.
If the proposals are accepted, then we set ${M_n} := {M}^\prime$, $\A_n := \A^\prime$, and $l_n := l^\prime$.
Otherwise, we set ${M_n} := {M}_{n-1}$, $\A_n := \A_{n-1}$, and $l_n := l_{n-1}$.}

\begin{algorithm}
	\caption{SpaRJ algorithm}
	\label{alg:gensparthreestep}
	\small
	\begin{algorithmic}
		\Input $\mathbf{y}_{1:T}, \mathbf{A}_0, g(\cdot), \pi_{0}, \pi_{-1}, N, \Lambda(\cdot,\cdot,\bm\lambda),\P_0,\Q,\R,\H,\bar{\mathbf{x}}_0$
		\Output Set of $N$ samples $\{\A_n, l_n, M_n\}_{n=1}^{N}$
		\State \textsl{\textbf{Initialisation}}
		\State Initialise $M_0$ as fully dense
		\State Evaluate filtering equations, obtaining $l_0 := \log(\mathrm{p}(\mathbf{y}_{1:T}|\A_0))$
		\For{$n =1,...,N$}
			\State Set $c := 0$.
			\State \textsl{\textbf{Step 1: Propose model}} (Section~\ref{sec:modjump})
			\State Run Algorithm~\ref{alg:multirules} to propose ${M}^\prime$
			\State \textsl{\textbf{Step 2: Propose $\A^\prime$}} (Section~\ref{sec:paramsampling})
			\State Run Algorithm~\ref{alg:parammap} to propose $\A^\prime$ and compute $c$
			\State \textsl{\textbf{Step 3: MH accept-reject}} (Section~\ref{sec:mhaccrej})
			\State Evaluate Kalman filter with $\A := \mathbf{A}^\prime$
			\State Set $l^\prime := \log(p(\mathbf{y}_{1:T}|\A^\prime))$
			\State Compute $\log(a_r) := l^\prime - l_{n-1} + \Lambda(\A_{n-1},\A^\prime,\bm\lambda) + c$
			\State \textit{Accept} w.p. $a_r$
			\If{\textit{Accept}}
				\State Set ${M_n} := {M}^\prime$, $\A_n := \A^\prime$, $l_n := \log(p(\mathbf{y}_{1:T}|\A^\prime))$
			\Else
				\State Set $M_n := M_{n-1}, \A_n := \A_{n-1}, l_n := l_{n-1}$
			\EndIf
		\EndFor
	\end{algorithmic}
\end{algorithm}

\begin{algorithm}
	\caption{Model proposal routine}
	\label{alg:multirules}
	\small
	\begin{algorithmic}
		\Input $M_{n-1}, \pi_{-1}, \lambda_j$
		\Output $M^\prime, {\{I_i\}_{i=1}^k}$
		\State \textbf{\textsl{Step 1: Determine jump}}
		\State \textit{Retain} w.p. $\pi_0$
		\If{\textit{Retain}}
			\State Set ${M}^\prime := {M}_{n-1}$
		\Else
		\State \textbf{\textsl{Step 1.1: Determine jump direction}}
		\If{$|\mathcal{M}_{n-1}| = 0$}
			\State \textit{Jump denser}
		\ElsIf{$|\mathcal{M}_{n-1}| = d_x^2$}
		    \State \textit{Jump sparser}
		\Else
			\State \textit{Jump sparser} w.p. $\pi_{-1}$, else \textit{jump denser}
		\EndIf
		\State \textsl{\textbf{Step 1.2: Perform jump}}
		\If{\textit{Jump sparser}}{ (Step 1.2s)}
			\State Draw $k \sim \text{TPoi}(\lambda_j, 1, D_n)$ (See App.~\ref{app:TPoi})
			\State Select $k$ elements {$\{I_i\}_{i=1}^k$} of $\mathcal{M}_{n-1}$
			\State Set $M^\prime$ such that $\mathcal{M}^\prime = \mathcal{M}_{n-1} \setminus {\{I_i\}_{i=1}^k}$
		\Else{ \textit{Jump denser} (Step 1.2d)}
			\State Draw $k \sim \text{TPoi}(\lambda_j, 1, S_n)$ (See App.~\ref{app:TPoi})
			\State Select $k$ elements ${\{I_i\}_{i=1}^k}$ of $\mathcal{M}_{n-1}^\stcomp$ 
			\State Set $M^\prime$ such that $\mathcal{M}^\prime = \mathcal{M}_{n-1} \cup {\{I_i\}_{i=1}^k}$
		\EndIf
		\EndIf
	\end{algorithmic}
\end{algorithm}

\begin{algorithm}
	\caption{Parameter proposal routine}
	\label{alg:parammap}
	\small
	\begin{algorithmic}
		\Input $\mathbf{A}_{n-1}, g(\cdot), M^\prime, {\{I_i\}_{i=1}^k}, c$
		\Output $\mathbf{A}^\prime, c$
		\State \textsl{\textbf{Step 2: Determine jump}}
		\If{Retain}
		\State \textsl{\textbf{Step 2.1: Sample posterior}}
		\State Propose $\mathbf{A}^\prime$ from the posterior under $M^\prime$, $p(\A|M^\prime,\y_{1:T})$.
		\Else
		\State \textsl{\textbf{Step 2.2: Map parameters}}
		\If{\textit{Jump sparser}}{ (Step 2.2s)}
		\State Set $\mathbf{A}^\prime := \mathbf{A}_{n-1}$ with elements ${\{I_i\}_{i=1}^k}$ set to $0$.
		\State Set $c := \sum_{i=1}^k\log(g(a_{I_i}))$.
		\Else { (Step 2.2d)}
		\State Draw $u_1, \dots, u_k \sim g(\cdot)$. 
		\State Set $\mathbf{A}^\prime := \mathbf{A}_{n-1}$ with elements ${\{I_i\}_{i=1}^k}$ set to $u_i$. 
		\State Set $c := -\sum_{i=1}^k\log(g(u_i))$.
		\EndIf
		\State Modify $c$ as per Appendix~\ref{app:corr}. 
		\EndIf
	\end{algorithmic}
\end{algorithm}

\section{Algorithm design}
\label{sec:sparjdisc}
We now detail the three steps of Algorithm~\ref{alg:gensparthreestep} as presented in Section~\ref{sec:sparj}.
This section is structured to follow the steps of the algorithm for clarity and reproducibility.

\subsection{Step 1: Model sampling}
\label{sec:modjump}
In order to explore potential sparsity of $\A$ using RJMCMC, we design a model jumping scheme that exploits the structure inherent to the model space.

\subsubsection{Model jumping scheme (steps 1.1 and 1.2 in Alg.~\ref{alg:multirules})}
\label{sec:jsoutline}

At each iteration, the algorithm proposes to jump models with probability (w.p.) $1-\pi_0$, as in Step 1 of Algorithm~\ref{alg:multirules}.
If the algorithm proposes a model jump, then the proposed model $M^\prime$ will be sparser than $M_{n-1}$ w.p. $\pi_{-1}$, and denser than $M_{n-1}$ otherwise.
If no model jump is proposed, then $M^\prime = M_{n-1}$.

{There are thus three distinct outcomes of the model jumping step: retention of the previous model, proposing to jump to a sparser model, or proposing to jump to to a denser model.
Note that in some cases it is not possible to jump in both directions, and hence if the jumping scheme proposes a model jump, the jump direction is deterministic.
This changes the model jump probability, with the results detailed in} Appendix~\ref{app:corr}.
\subsubsection{Model space adjacency (steps 1.2s and 1.2d in Alg.~\ref{alg:multirules})}
\label{sec:modadj}
Given the jump direction from Step 1.1, we denote by $k$ the number of elements that are to be made sparse or dense.
We draw $k \sim \text{TPoi}(\lambda_j, 1, m_n)$ (see Appendix~\ref{app:TPoi}), where $m_n$ is the maximum jump distance in the chosen direction, equal to $S_n$ if jumping denser, and $D_n$ if jumping sparser.
The rate $\lambda_j$ should be chosen with $\lambda_j \in [0,1)$ to prefer jumps to closely related models. 
We find experimentally that $\lambda_j = 0.1$ gives good results, and note that $\lambda_j = 0$ is equivalent to a scheme in which the sparsity can change by one element only. {Due to the small size of the space in $\lambda_j$ a grid search would also be possible}. 
The resulting model jump probabilities are asymmetric, with the modification to the acceptance ratio given in Appendix~\ref{app:corr}.

In order to provide a set of candidate models for $M^\prime$, we impose an adjacency condition.
We say model $M^{(1)}$ is densely $k$-adjacent to model $M^{(2)}$ if both $|\mathcal{M}^{(1)} \backslash \mathcal{M}^{(2)}| = k$ and $|\mathcal{M}^{(2)} \backslash \mathcal{M}^{(1)}| = 0$, and sparsely $k$-adjacent if both $|\mathcal{M}^{(2)} \backslash \mathcal{M}^{(1)}| = k$ and $|\mathcal{M}^{(1)} \backslash \mathcal{M}^{(2)}| = 0$.
In other words, if the model $M^{(1)}$ differs in $k$ elements from $M^{(2)}$ in a given direction only, then it is $k$-adjacent in that direction; if the model $M^{(1)}$ differs from $M^{(2)}$ in both the sparse and dense directions, e.g. two elements are sparser and one element is denser, then it is not adjacent to $M^{(2)}$. 
Note that, if $M^{(1)}$ is sparsely $k$-adjacent to $M^{(2)}$, then $M^{(2)}$ is densely $k$-adjacent to $M^{(1)}$, satisfying the reversibility condition of RJMCMC.
With this adjacency condition, for a given $k$ and jump direction, the proposal $\M^\prime$ is uniformly selected from the models $k$-adjacent to $M_{n-1}$ in the given direction.
Note that, given $\lambda_j > 0$, it is theoretically possible to reach any model in a maximum of two jumps: one to the maximally sparse model and one jump denser to the desired model.

It is possible to extend this proposal technique to jumping in both direction simultaneously, rather than requiring combinations of birth-death moves to achieve the result.
This is omitted for simplicity.
The natural solution to this is evaluating each jump with an accept-reject step, which effectively recovers the current scheme.
It is also possible to propose $\mathcal{M}^\prime$ as a randomly selected list of indices with length $D_n$, effectively shuffling the sparse elements around.
This could potentially allow for more robust exploration of the posterior, but these moves would be very unlikely to be accepted.
We therefore believe our proposal method to be a good compromise between simplicity and robustness, with good performance as evidenced by Section~\ref{sec:numerical}.

\subsubsection{Choice of parameters for model jumps}
The value of the hyper-parameters $\pi_0$ and $\pi_{-1}$ affects the acceptance rate of the proposed models and parameter values.
It is known that an acceptance rate close to $0.234$ is optimal for a random walk Metropolis-Hastings sampler \cite{gelman1997weak}, and works well as a rule of thumb for RJMCMC algorithms \cite{gagnon2019weak}.
We aim to have our within-model samples accepted at close to this rate, and thus must not propose to change model too often.
This is because model changes can significantly alter the conditional posterior $p(\A|M^\prime)$, often leading to a low acceptance probability.
We recommend using a model retention probability of $\pi_0 \approx 0.8$. 
We find this gives enough iterations per model to average close to the optimal acceptance rate, whilst also proposing to jump models relatively frequently, allowing for exploration of sparsity. 
We recommend setting $\pi_{-1} = 0.5$, making the model proposal process symmetric, although $\pi_{-1}$ can reflect prior knowledge of the sparsity of $\A$, with larger $\pi_{-1}$ indicating a preference for sparsity. 
{The algorithm is relatively insensitive to the value of $\pi_0$ and $\pi_{-1}$, allowing for these parameters to be chosen easily.}
However, $\pi_0$ and $\pi_{-1}$ can be tuned during the burn-in period, with the objective of reaching a given acceptance rate.
{We find that using a value of $0.5$ works well for both parameters, due to the structure of the model space and restricted parameter space of stable LGSSMs}

\subsection{Step 2: Parameter sampling and mapping}
\label{sec:paramsampling}
{Since our method applies MCMC to sample the posterior distribution $p(\A|\y_{1:T})$, we must define a parameter proposal routine.}
Once a model has been proposed, the algorithm proposes a parameter value, $\A^\prime$, which is constrained to the parameter space of $M^\prime$. 
The process by which we generate the parameter proposal depends on whether or not the proposed model $M^\prime$ is the same as the previous model $M_{n-1}$.
If $M^\prime = M_{n-1}$, the conditional posterior of the transition matrix, $p(\A|\y_{1:T}, M^\prime)$, is sampled. 
Otherwise, the parameter value is mapped to the parameter space of $M^\prime$. 

\subsubsection{Sampling $\A$ under a given model (Step 2.1)}
\label{sec:rwmhsec}
To generate the parameter proposal $\A^\prime$, we sample from $p(\A|M^\prime,\y_{1:T})$, the posterior distribution of the transition matrix $\A$ under the model $M^\prime$.
This distribution can be written as
\begin{equation}
	\begin{aligned}
		p(\A|M^\prime,\y_{1:T}) &= \int p(\x_{0:T},\A|\y_{1:T}, M^\prime) \ \mathrm{d}\x_{0:T}\\
		&\propto p(\A|M^\prime)p(\y_{1:T}|\A),
	\end{aligned}
\end{equation}
where $p(\A|M^\prime)$ is the prior assigned to the transition matrix under the proposed model, which can be written
\begin{equation}
	\label{eq:aprior}
	p(A_{ij}|M_n) \sim \begin{cases}
		p(A_{ij}), & (i,j) \in \mathcal{M}^\prime,\\
		\delta_0, & \text{otherwise},\\
	\end{cases}
\end{equation}
with $p(A_{ij})$ deriving from $p(\A)$, and $\delta_0$ denoting a point mass at $0$. By substituting $\A$ into eq.~\eqref{eq:ylikdecom} we obtain
\begin{equation}
	p(\y_{1:T}|\A) = p(y_1|\A)\prod_{i=1}^{T}p(\y_i|\y_{1:i-1},\A),
\end{equation}
{and can hence evaluate $p(\A|M^\prime)p(\y_{1:T}|\A)$, allowing us to sample from $p(\A|M^\prime,\y_{1:T})$.}

We propose to sample from this distribution using a random walk Metropolis-Hastings (RWMH) sampler. 
For the walk distribution, we use a Laplace distribution for each element of $\A$, with all steps element-wise distributed i.i.d. $\text{Laplace}(0,\sigma)$, {with $\sigma$ discussed below.}
We can thus view our {proposed} parameter proposal as drawing from 
\begin{equation}
	\label{eq:aprop}
	(A^\prime)_{ij} \sim \begin{cases}
		\text{Laplace}((A_{n-1})_{ij}, \sigma), & (i,j) \in \mathcal{M}^\prime,\\
		\delta_0, & \text{otherwise}.\\
	\end{cases}
\end{equation}
{The Laplace distribution is selected, primarily due to its relationship with our proposed prior distribution for $\A$, itself a Laplace distribution \cite{park2008bayesian}.}
In addition, the mass concentration of the Laplace distribution means that the walk will primarily propose values close to the previous value, increasing the acceptance rate, but can also propose values that are further from the accepted value, improving the mixing of the sample chain. 
The value of $\sigma$ is chosen to give a within-model acceptance rate near the optimal rate of $0.234$ \cite{gelman1997weak}, with $\sigma = 0.1$ consistently yielding rates close to this. 
{A grid search suffices to select the value of this parameter as, for stable systems, the space of feasible values is small ($\sigma < 1$ is recommended).}

\subsubsection{Completion distributions (steps 2.2s, 2.2d)}
\label{sec:compl}

In our algorithm, when a model jump occurs the dimension of the parameter space always changes.
For example, jumping to a sparser model is equivalent in the parameters space to discarding parameters and decreasing the dimension of the parameter space.
However, if jumping to a denser model, hence increasing the dimension of the parameter space, we require a method to assign a value to the new parameter.
RJMCMC accomplishes this by augmenting the parameter mapping from model $M^{(i)}$ to $M^{(j)}$ with draws from a completion distribution $g_{i,j}(\cdot)$, defined for each possible model jump \cite{green1995reversible}.

Rather than defining a distribution for every pair of models, we exploit the numerical properties of sparsity to define a global completion distribution $g(\cdot)$. 
In our samples, sparse elements take the value zero.
In order to propose parameters close to the previous parameters, we draw the value of newly dense elements such that the value is close to zero.
In order to accomplish this, we choose a $\text{Laplace}(0,\sigma_c)$ as the global completion distribution $g(\cdot)$, due to its mass concentration and relation to the prior we propose in Section~\ref{sec:priors}. 
The $\sigma_c$ parameter is subject to choice, and for stable systems we find that $\sigma_c \approx 0.1$ performs well, although the parameter could be tuned during the burn-in period. 
{A simple grid search suffices to select the value of this parameter as the method is robust to parameter specification via the accept-reject step \cite{robert2013monte}.
For stable systems, the search space for this parameter is small, approximately $(0, 0.5]$, so a grid search is most efficient in terms of computational cost.}
If the prior is chosen as per the recommendations in Section~\ref{sec:mhaccrej}, then we can interpret this renewal process as drawing the values for newly dense elements from the prior.
\subsubsection{Mapping between parameter spaces (steps 2.2s, 2.2d)}
\label{sec:bijec}
In order to jump models, we must be able to map between the parameter spaces of the previous model and the proposal.
This is, in general, a difficult task \cite{green1995reversible, robert2013monte}, but is eased in our case as the models we are sampling are specific cases of the same model, and thus the parameters are the same between models.
We therefore use an augmented identity mapping to preserve the interpretation of parameter values between models.
Written in terms of $\A$, this mapping is given by
\begin{align}
	\begin{split}
		(A^\prime)_{ij} &= 
		\begin{cases}
			(A_{n-1})_{ij}, & \text{if $A_{ij}$ is unchanged},\\
			u_{ij}, & \text{if $A_{ij}$ becomes dense},\\
			0, & \text{if $A_{ij}$ becomes sparse},\\
		\end{cases}\\
	\end{split}
\end{align}
with $u_{ij}$ i.i.d. $g(\cdot)$. 
In order to obtain the Jacobian required to evaluate eq.~\eqref{eq:rjaccrat}, the transformation must be written as applied to the parameter space, giving
{
\begin{equation}
	\label{eq:tfj}
	(A^\prime)_{ij} = 
	\begin{cases}
		(A_{n-1})_{ij}^{(n-1)}, & (i,j) \in \mathcal{M}_{n-1} \cap \mathcal{M}^\prime,\\
		u_{ij}, & (i,j) \in \mathcal{M}^\prime \setminus \mathcal{M}_{n-1}.\\
	\end{cases}
\end{equation}}
As sparse elements are, by construction, not in the parameter space, they are not present in the transformation, and are taken to be zero in $\A^\prime$ by definition. 
The Jacobian of the parameter mapping given by eq.~\eqref{eq:tfj} is a $D^\prime \times D^\prime$ identity matrix, and hence the Jacobian determinant term in Eq.~\eqref{eq:rjaccrat} is constant and equal to one.

\subsection{Step 3: MH accept-reject}

In this section, we first discuss the MH acceptance ratio, and then the way that prior knowledge of the transition matrix is incorporated, and how this relates to the model space.
We then discuss the implications of sparsity in the samples.

\subsubsection{Modified acceptance ratio}
\label{sec:mhaccrej}
The modified Metropolis-Hastings acceptance ratio for our method is given by 
\begin{equation}
	a_r^{(n-1,n)} = \frac{g(u_{n})}{g(u_{n-1})}\frac{\pi_{n,n-1}}{\pi_{n-1,n}}\frac{p(\A^\prime)p(\y_{1:T}|\A^\prime)}{p(\A_{n-1})p(\y_{1:T}|\A_{n-1})}.
	\label{eq:rjaccrat2}
\end{equation}
The first term is a correction for detailed balance, required due to the stochastic completion of the parameter mappings.
The second term is analogous to the modification required when using an asymmetric proposal in RWMH, but relating to the model space. 
The last term of the expression is the standard symmetric Metropolis acceptance ratio. 
Note that the Jacobian term from eq.~\eqref{eq:rjaccrat} is equal to one in our case, and is therefore omitted.

\subsubsection{Incorporating prior knowledge}
\label{sec:priors}   
The prior distribution, $p(\A)$, quantifies our pre-existing knowledge on $\A$, including both its sparsity structure.
We do not enforce sparsity via this distribution. 
Since it encodes our knowledge of all elements of $\A$ we call it the overall prior.
We can interpret the prior of $\A$ conditional on a given model $M_n$, or $p(\A|M_n)$, as encoding the sparsity from the model, and write it as in eq.~\eqref{eq:aprior}.
In this way, we can interpret the sparsity constraint as a series of priors. 
The relationship between the prior conditional on the model and the overall prior is given by 
\begin{equation}
\label{eq:priorassum}
	p(\A) = \sum_{M} p(\A|M)p(M),
\end{equation}
where $p(\A)$ is the overall prior, and $p(\A|M)$ the conditional prior containing sparsity in elements indexed by $\mathcal{M}$, given in eq.~\eqref{eq:aprior}, and $p(M)$ is the prior assigned to the model space, which is described in Section~\ref{sec:modadj}.

{
We apply the prior via the function $\Lambda(\A_{n-1},\A^\prime,\bm\lambda)$, where $\bm\lambda$ is a vector of prior hyper-parameters. When written in terms of the prior, $$\Lambda(\A_{n-1},\A^\prime,\bm\lambda) = \log(p(\A^\prime;\bm\lambda)) - \log(p(\A_{n-1};\bm\lambda)).$$}
{We recommend an element-wise Laplace prior on the transition matrix $\A$, given by $p(A_{ij}) := \text{Laplace}(0, \lambda)$ with $\lambda$ subject to choice, which results in  
\begin{equation}
\label{eq:lassolog}
	\Lambda(\A_{n-1}, \A^\prime, \lambda) = \lambda(\lVert \mathbf{A}_{n-1} \rVert_1^1 - \lVert \mathbf{A}^\prime \rVert_1^1)
\end{equation}
after combining all $p(A_{ij})$ to yield $p(\A)$.
This is equivalent to the LASSO penalty \cite{tibshirani1996regression, park2008bayesian} in regression, which is known to promote sparsity.
Experimentally, we find that choosing $\lambda \in [\exp(-2), \exp(2)]$ consistently results in good performance.
For more information on selecting parameters for the Laplace prior, and the Laplace prior in general, we refer the reader to \cite{park2008bayesian}.
{Note that the Laplace prior is the Bayesian equivalent to LASSO regression \cite{tibshirani1996regression, park2008bayesian}, with penalties and priors having an equivalence in Bayesian statistics, as both encode the prior knowledge of a parameter.}
}

If the parameter $\lambda$ is not determined based on prior knowledge, it is possible to use a simple grid search to select the value of the parameter as explained above.
Furthermore, if a $\text{Laplace}(0,\sigma_c)$ is used for the completion distribution $g(\cdot)$, then dense values are re-initialised using the prior, reinforcing the interpretation of the prior as our existing knowledge.
Note that using this prior is not required to recover sparsity, as this occurs as a result of the model sampling.
We have performed several runs utilising a diffuse prior on the parameter space, with the results being similar to those where our suggested prior is used.
{Priors other than Laplace can be used, without compromising the recovery of sparsity, such as the ridge prior, or a diffuse prior.}
We use the LASSO penalty for its connection to sparsity, as well as for direct comparability to the existing literature.

As the parameter proposal uses a standard MCMC scheme, a prior sensitivity analysis can be used to determine the effect of the chosen prior.
{In order to validate our recommendations, we have run multiple sensitivity analyses for the diffuse prior and several Laplace priors of varying scale, and have observed that the results are independent of the prior in all but the most extreme cases in which the prior is nearly a point mass.}
\subsubsection{Model space prior}
\label{sec:modprior}
As the model space encodes only the sparsity of $\A$, a prior on $\A$ that incorporates this structure is also implicitly a prior on the model space.
{If no such prior is applied, then the implicit prior on the model space is diffuse, with $p(M) = (2^{-d_x^2}) \ \forall M$.}
This follows from evaluating $p(\A)$ at an arbitrary $\A$ under each model via Eq.~\eqref{eq:priorassum}, giving $p(\A) = p(\A|M)$ when $\A$ has the sparsity structure induced by $M$, noting $p(\A|M) = 0$ if $\A$ does not have this structure.
{As this holds for all models, it follows that $p(M) \propto 1$, and as the model space is discrete and finite, we can obtain an explicit value for the prior.}
{Note that a diffuse prior is, in general, not allowed on the model space if using a posteriori model comparison methods \cite{llorente2022safe}. However a diffuse prior is standard for RJMCMC \cite{green1995reversible, robert2013monte, richardson1997bayesian}, as the model space is sampled, and the model dynamically assessed alongside the parameter.}
{This diffuse prior encodes our lack of prior knowledge as to the specific sparsity structure of $\A$.}

\subsubsection{Probabilistic Granger causality}
In LGSSMs, an element $x_i$ of the state space Granger-causes element $x_j$ if knowledge of $x_i$ at time $t$ improves the prediction of $x_j$ at time $t+1$.
We can therefore derive probabilistic Granger-causal relationships from our samples of the transition matrix, as the sampler assesses the proposals using their likelihood, which is equivalent to assessing their predictive capabilities. 
These probabilistic Granger-causal relationships are powerful, as they allow the probability of a relationship between variables to be quantified.

{Note that $A_{ji}$ being zero in the transition matrix does not necessarily mean independence of the state elements, but directed conditional independence on the scale of one time step.}
{This conditional independence means that $x_i$ does not Granger-cause $x_j$, however $x_i$ may indirectly affect $x_j$ through another variable over multiple time steps.}
\subsection{{Extending SpaRJ to other parameters}}
\label{sec:otherparams}
{
Our method can be used to obtain sparse estimates of any of the parameters of the LGSSM, although some modifications are required to extend the formulation given above (for the transition matrix).
Extending our method to the matrix $\H$ requires only for the parameter and model proposal to be changed to reflect the size of $\H$. }

{
Extending the method to covariance matrices requires the proposal value to be constrained such that the resulting matrix is positive semi-definite. 
If the method does not jump models, remaining in model $M_{t-1}$, a possible {proposal} distribution for the state covariance proposal $\Q'$ is }
\begin{equation}
    {\Q' \sim \mathrm{Wishart}(\Q_{t-1}/p, p),}
\end{equation}
{
where $p$ is larger than $d_x$, with $p > 30d_x$ working well experimentally. 
This proposal has expectation close to $\Q_{t-1}$, and is positive semi-definite. 
{Note that this distribution is not applied to $\Q$ if estimating only $\A$, e.g., in Section~\ref{sec:sparj} or in other subsections of Section~\ref{sec:sparjdisc}, and applies only to the extension to sampling the covariance parameter.}
We enforce the sparsity structure of $\Q_{t-1}$ in $\Q'$, by setting elements sparse under $M_{t-1}$ to $0$ after sampling but before the accept-reject step, replacing Step 2.1 in Algorithm~\ref{alg:parammap}, where the model does not change. 
}

{
The model proposal step (Step 2.2 in Algorithm~\ref{alg:parammap}) would also need to be modified, with the diagonal being dense at all times, and enforcing indices $(i,j)$ and $(j,i)$ to have the same sparsity. 
The model space is therefore reduced, and model adjacency is assessed via only the upper triangular.
This modification to the model proposal process completes the alterations required to use our method to sparsely sample the state covariance matrix.
Note that the covariance parameters cannot be interpreted as encoding state connections, and therefore cannot be interpreted graphically in the same way as in the transition matrix.
}

\subsection{{Computational cost}}
{
The computational cost of our method is very similar to that of regular MCMC methods when applied to state-space models. 
The most computationally intensive component of the algorithm is the evaluation of the Kalman filtering equations, with this being over $95\%$ of the computational time in our testing.
The additional costs compared to a {random walk Metropolis-Hastings (RWMH)} method are one or two draws from a uniform distribution, zero or one draw from a truncated Poisson distribution (equivalent to a categorical distribution), and some additional array accesses and comparisons.
These extra costs are negligible compared to the cost of evaluating the Kalman filtering equations, with the computational cost and complexity being determined by the matrix operations therein, {resulting in a complexity of $O(NT(d_x^3 + d_y^3))$ for our algorithm, where the dominating $d_x^3$ and $d_y^3$ terms result from the matrix operations performed by Kalman filter. }
The computational cost of our method is empirically demonstrated in Section~\ref{sec:numerical}, and is functionally equivalent to a standard MCMC method that does not explore sparsity. 
Thus in practice, given the cost of the filtering equations, the sparsity is explored for free.}

\label{sec:extend}

\section{Numerical study}
\label{sec:numerical}

We now present the results of three sets of simulation studies to evaluate our method, showcasing the performance of SpaRJ in several scenarios.
The section is divided into three synthetic data experiments and one real data problem.
{First, we evaluate the method with isotropic covariance matrices over variable $d_x$ and $T$.
Next, we investigate the effect of known and unknown anisotropic state covariance $\Q$ over variable $d_x$ and $\lambda$. 
The third experiment explores the effect of the true level of sparsity $D$ in the transition matrix on the quality of inference. 
We then use real data to recover geographical relationships from global temperature data.
Finally, we explore the convergence characteristics of the method and check guarantees.}

For the synthetic experiments, we generate observations following eq.~\eqref{eq:lgssm}, with $d_x = d_y$, $\H = \Id_{d_x}$, and take $\bar{\x}_0 = \mathbf{1}_{d_x}$ and $T = 100$ unless stated otherwise.
The state covariance matrix $\Q$ is specified per study.
{We generate transition matrices and synthetic data for $d_x \in \{3, 6, 12\}$. 
Whilst this may seem limited in dimension, this equates to performing inference in $9, 36,$ and $144$ dimensional spaces, as each element of the transition matrix is an independent parameter.
Furthermore, we sample the model space, which is of size $2^{d_x^2}$, e.g. $2^{144}$ when $d_x = 12$.}
In all experiments, we run SpaRJ for $N=15000$ iterations, discarding the first $5000$ as burn-in.
The matrix $\A_0$ is generated using an EM scheme, initialised at a random element-wise standard normal matrix.
We set $\pi_0 = 0.8$ and $\pi_{-1} = 0.5$ in all cases.
The LASSO penalty is used, with $\lambda$ chosen per experiment. 
We use a truncated Poisson distribution for the jump size, with $\lambda_j = 0.1$. 

We contrast our proposed method method with GraphEM \cite{chouzenoux2020graphem,elvira2022graphical}, an algorithm with similar goals based on proximal optimisation. 
In addition, we compare with the conditional Granger causality (CGC) method of \cite{luengo2019hierarchical} {and the DAG-based method (DAGMA) of \cite{bello2022dagma}. 
These methods do not exploit the state-space model structure, and are trained only on the observations $\y_{1:T}$.}
{We note that there are not other RJMCMC-based methods that are applicable to this problem.}
We therefore compare with a reference MCMC implementation that does not exploit sparsity, and is hence dense in all elements of the estimate. 
{This is equivalent to running our method, but with $p(M) = 0$ except for the $M$ corresponding to the fully dense $\A$ matrix, hence effectively removing Step 1 in Algorithm~\ref{alg:gensparthreestep} and Section~\ref{sec:sparj}.}
We compare the metrics of RMSE, precision, recall, specificity {(true negative rate)}, and F1 score, {with an element being sparse encoded as a positive, and dense as a negative. }

{We use these metrics, which are associated with classification rather than regression, as our method outputs truly sparse samples, and therefore allows for parameters to be classified as sparse or dense without thresholding on their numerical value, or using confidence/credible intervals.}
{We take an element to be sparse under SpaRJ by majority vote of the samples. }
We average the metrics over 100 independent runs of each algorithm. 
{The average time taken runs to complete is given, with the runs being performed in parallel on an 8 core processor. No special effort was put into optimising any single method, and all methods that use the Kalman filter utilise the same implementation thereof. Note that the DAGMA implementation is GPU accelerated, whereas all other methods utilise only the CPU.}
RMSE for SpaRJ and for the reference MCMC is calculated with respect to the mean of post-burnin samples for each chain. {Note the RMSE is computed relative to $\A$, not the sequence of underlying hidden states as is often the case.}
{RMSE is not meaningful for CGC, as it estimates only connectivity.}
We generate our $\A$ matrices by drawing the dense elements from a standard normal, and then divide $\A$ by the magnitude of its maximal singular value to give a stable system.

\subsection{Synthetic data validation}
\label{sec:numknown}

\subsubsection{Isotropic covariances $\Q$ and $\R$}
\label{sec:numiso}
We test the performance of the method with isotropic covariance matrices $\Q$ and $\R$.

\noindent{\textbf{Dimension 3 matrix.}}  
{We generate $\A$ for dimension $d_x=3$ with sparsity in one element per row and one element per column. We set $\Q = \R = \Id_{d_x}, \P = 10^{-8}\Id_{d_x}$, and $\lambda = 1 = \exp(0)$.}

\noindent{\textbf{Dimension 6 block diagonal matrix.}}  
{We generate $\A$ for dimension $d_x=6$ as a block diagonal matrix with $2\times2$ blocks. We set $\Q = \R = 10^{-2}\Id_{d_x}$, $\P = 10^{-8}\Id_{d_x}$, and $\lambda = \exp(-1) \approx 0.367$. }

\noindent{\textbf{Dimension 12 block diagonal matrix.}}  
{We generate $\A$ for dimension $d_x=12$ as a block diagonal matrix with $2\times2$ blocks. We set $\Q = \R = 10^{-2}\Id_{d_x}$, $\P = 10^{-8}\Id_{d_x}$, and $\lambda = \exp(-1) \approx 0.367$. } 
\begin{table}[ht]
	\centering
	\footnotesize 
	\caption{Results for systems with known isotropic state covariance $\Q$, alongside average time per run. }
	\begin{tabular}{|c||c|c|c|c|c|c|c|}
		\hline
		$d_x$ & method & RMSE & spec. & recall & prec. & F1 & {Time (s)}\\
		\hline
		\hline
		$3$ & GraphEM & 0.099 & 0.86 & 0.98 & 0.79 & 0.88 & \textbf{0.043}\\
		& SpaRJ & \textbf{0.092} & \textbf{0.98} & \textbf{0.99}& \textbf{0.99} & \textbf{0.99} & 0.68\\
		& CGC & -- & 0.85 & 0.95 & 0.87 & 0.92 & 0.32\\
		& {DAGMA} & 0.49 & 0.17 & 0.67 & 0.29 & 0.40 & 0.25\\
		& MCMC & 0.103 & 1 & 0 & -- & 0 & 0.65\\
		\hline
		$6$ & GraphEM & 0.103 & 0.83 & 0.90 & 0.91 & 0.91 & \textbf{0.22}\\
		& SpaRJ & \textbf{0.094} & \textbf{0.88} & \textbf{0.96} & \textbf{0.94} & \textbf{0.95} & 3.2\\
		& CGC & -- & 0.87 & 0.93 & 0.91 & 0.90 & 1.6\\
		& {DAGMA} & 0.27 & 0.17 & 0.96 & 0.69 & 0.87 & 0.62\\
		& MCMC & 0.114 & 1 & 0 & -- & 0 & 3.3\\
		\hline
		$12$ & GraphEM & 0.090 & 0.85 & 0.77 & \textbf{0.96} & 0.85 & \textbf{0.93}\\
		& SpaRJ & \textbf{0.071} & 0.83 & 0.89 & 0.91 & 0.90 & 14.5\\
		& CGC & -- & 0.80 & 0.67 & 0.75 & 0.71 & 6.5\\
		& {DAGMA} & 0.16 & 0.25 & \textbf{0.98} & 0.86 & \textbf{0.92} & 1.0\\
		& MCMC & 0.107 & 1 & 0 & -- & 0 & 14.4\\
		\hline
	\end{tabular}
	
	\label{tab:systemtypes}
\end{table}

Table~\ref{tab:systemtypes} evidences a good performance from SpaRJ, exhibiting the capability to extract the sparsity structure in all examples.
Furthermore, point estimates resulting from SpaRJ are consistently closer to the true value than those from comparable methods, {as evidenced by the lower RMSE}.
{Note that in all cases the DAG based method recovered overly sparse graphs, as evidenced by the poor specificity scores. 
We further note that DAGMA is designed to recover acyclic graphs, with all graphs here being cyclical, further degrading performance.}

\begin{figure}[H]
	\centering
	\includegraphics[width=0.45\textwidth]{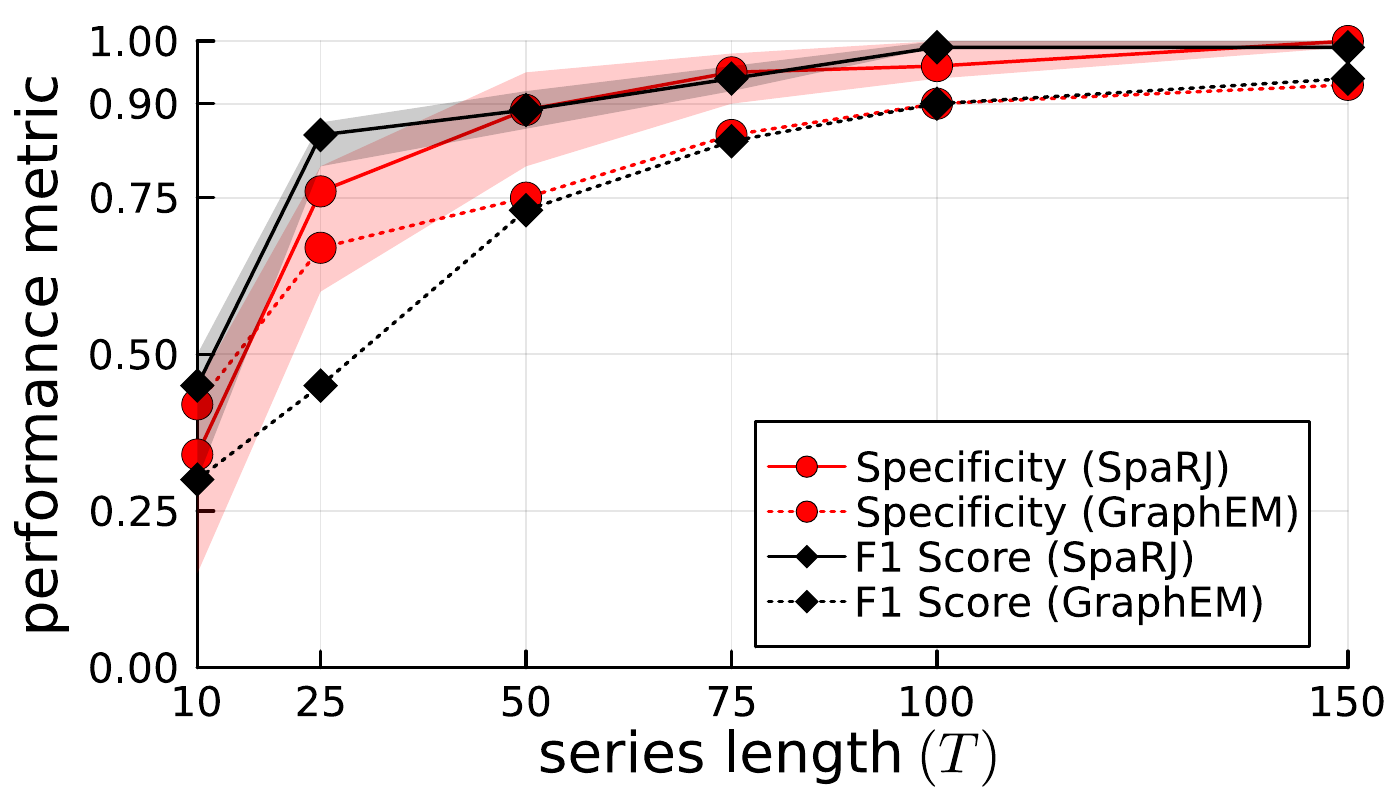}
	\caption{Sparsity metrics for variable series length $T$ for a $3\times3$ system with known isotropic state covariance. Shaded regions denote $95\%$ HPDIs (highest posterior density intervals), markers denote means. The dotted line indicates the mean performance of GraphEM.}
	\label{fig:serieslength}
\end{figure}

{In order to test the relationship between the recovered values and the number of observations $T$, we now demonstrate our method for different values of $T \in [10,150]$ using the same $3\times3$ system as previously. 
In Figure~\ref{fig:serieslength}, we show averaged metrics over 100 independent runs for SpaRJ and GraphEM. 
We see that the longer the series the better the overall performance, with SpaRJ giving a better overall performance than GraphEM}.

{The change in the quality of inference with the times series length $T$ illustrated in Figure~\ref{fig:serieslength} is typical for parameter estimation methods in state-space modelling, as a longer series gives more statistical information with which to perform inference.}

\subsubsection{Known anisotropic state covariance $\Q$}
\label{sec:numnoiso}
{We now generate synthetic data using a less favourable regime, under an anisotropic state covariance $\Q$.}
In order to do this, we note that all $n \times n$ covariance matrices $\bm\Sigma$ can be expressed in the form 
\begin{equation}
	\label{eq:randcov}\mathbf\Sigma = \G^{T}\text{Diag}(e_1, e_2, \dots, e_n)\G, 
\end{equation} 
where $\G$ is an orthogonal matrix, and $e_k$ are the eigenvalues of $\mathbf\Sigma$, with $e_1 \geq e_2 \geq \cdots \geq e_n > 0$.

To generate a covariance matrix, we first generate an orthogonal matrix $G$ following the algorithm of \cite{mezzadri2007generate}.
We then draw $e_i \sim \text{U}(0.5,1.5)$, and sort in descending order. 
Finally, we obtain $\mathbf\Sigma$ via evaluation of Eq.~\eqref{eq:randcov}. 
{In this way, we generate a random positive definite matrix, with all elements non-zero.}

{To allow direct comparison with the previous results, we use the same set of model parameters as before, except we randomly generate the covariance $\Q$ for each system as above.}

\begin{table}[ht]
	\centering
	\footnotesize
	\caption{Results for systems with known anisotropic state covariance $\Q$, alongside average time per run.}
	\begin{tabular}{|c||c|c|c|c|c|c|c|}
		\hline
		$d_x$ & method & RMSE & spec. & recall & prec. & F1 & {Time (s)}\\
		\hline
		\hline
		$3$ & GraphEM & \textbf{0.093} & 0.98 & 0.85 & 0.97 & 0.88 & \textbf{0.052}\\
		& SpaRJ & 0.087 & \textbf{0.98} & \textbf{0.99} & \textbf{0.98} & \textbf{0.99} & 0.67\\
		& CGC & -- & 0.72 & 0.93 & 0.43 & 0.59 & 0.32\\
		& {DAGMA} & 0.61 & 0.17 & 0.33 & 0.17 & 0.22 & 0.25\\
		& MCMC & 0.107 & 1 & 0 & -- & 0 & 0.68\\
		\hline
		$6$ & GraphEM & 0.09 & \textbf{0.99} & 0.48 & \textbf{0.98} & 0.63 & \textbf{0.24}\\
		& SpaRJ & \textbf{0.07} & 0.88 & 0.90 & 0.94 & \textbf{0.92} & 3.3\\
		& CGC & -- & 0.63 & 0.65 & 0.32 & 0.45 & 1.7\\
		& {DAGMA} & 0.26 & 0.25 & \textbf{0.96} & 0.71 & 0.82 & 0.62\\
		& MCMC & 0.09 & 1 & 0 & -- & 0 & 3.3\\
		\hline
		$12$ & GraphEM & 0.097 & \textbf{0.98} & 0.30 & \textbf{0.99} & 0.46 & \textbf{0.95}\\
		& SpaRJ & \textbf{0.082} & 0.95 & 0.83 & 0.99 & 0.90 & 14.4\\
		& CGC & -- & 0.75 & 0.57 & 0.43 & 0.49 & 6.5\\
		& {DAGMA} & 0.16 & 0.25 & \textbf{0.97} & 0.86 & \textbf{0.91} & 1.0\\
		& MCMC & 0.099 & 1 & 0 & -- & 0 & 14.4\\
		\hline
	\end{tabular}
	\label{tab:systemtypesniso}
\end{table}
{In Table~\ref{tab:systemtypesniso} see that there is only a small apparent difference in performance between isotropic covariance and non-isotropic state covariances, providing that the covariance is known. }
{This is expected, as a known covariance would not affect the estimation of the value of the state transition matrix.
However, when estimating sparsity, the anisotropic nature of the state covariance does have an effect. 
This is due to the value of the state elements affecting each other in more than one way, as is the case in the isotropic covariance case.
There is thus a small drop in metrics in all cases due to this additional source of error.
{We note that whilst DAGMA may seem to perform well due to the high F1 scores, it does this by recovering an overly sparse graph as indicated by the low specificity. For example, only 9 elements are recovered as dense in the $12$ dimensional system, out of a true 24 dense elements, which does not well represent the underlying system.}
}

{We now perform a sensitivity analysis, in which we vary the strength of the prior by varying $\lambda$, and observe the effect on the results. 
We will perform this analysis on the $d_x = 12$ system with a known anisotropic covariance.
The results of this analysis are presented in Table~\ref{tab:priorsense}. 
We see that the results are not dependent on the prior parameter, meaning that the parameter can be chosen without excess computation or prior knowledge required.}
\begin{table}[ht]
	\centering
	\footnotesize
	\caption{Results for variable penalty in the $d_x = 12$ known anisotropic system, alongside average time per run.}
	\begin{tabular}{|c||c|c|c|c|c|c|c|}
		\hline
		$\lambda$ & RMSE & spec. & recall & prec. & F1 & {Time (s)}\\
		\hline
		\hline
		$\exp(-1)$ & {0.082} & \textbf{0.95} & \textbf{0.83} & \textbf{0.99} & \textbf{0.90} & 14.2\\
		\hline
		$\exp(0)$ & {0.085} & 0.95 & 0.83 & 0.98 & 0.90 & \textbf{14.1}\\
		\hline
		$\exp(1)$ & {0.081} & 0.94 & 0.83 & 0.99 & 0.89 & 14.3\\
		\hline
		$\exp(-1.5)$ & {0.083} & 0.93 & 0.82 & 0.99 & 0.90 & 14.3\\
		\hline
		$\exp(-2)$ & \textbf{0.081} & 0.94 & 0.82 & 0.99 & 0.89 & 14.2\\
		\hline
		$0$ & 0.082 & 0.94 & 0.82 & 0.99 & 0.90 & 14.2\\
		\hline
	\end{tabular}
	\label{tab:priorsense}
\end{table}

\subsubsection{Estimated unknown anisotropic covariance}
\label{sec:numunknown}
In many scenarios, the true value of the state covariance $\Q$ is unknown, and must be estimated. 
{As we wish to assess the performance of our method in this scenario, we use the same true state covariance as Section~\ref{sec:numnoiso}, but input an estimated state covariance.
However, as both $\A$ and $\Q$ are now unknown, we must estimate both parameters in order to obtain an estimate for $\Q$. 
We therefore iteratively estimate $\A$ and $\Q$ using their analytic maximisers, and input the resulting estimate for $\Q$ into the tested methods.
The estimated $\A$ resulting from this initialisation is discarded, and is not used in our method, nor in any other method.}
\begin{table}[ht]
	\centering
	\footnotesize
	\caption{Results for systems with estimated anisotropic state covariance $\Q$, alongside average time per run.}
	\begin{tabular}{|c||c|c|c|c|c|c|c|}
		\hline
		$d_x$ & method & RMSE & spec. & recall & prec. & F1 & {Time (s)}\\
		\hline
		\hline
		$3$ & GraphEM & 0.123 & 0.75 & 0.72 & 0.62 & 0.65 & \textbf{0.05}\\
		& SpaRJ & \textbf{0.099} & \textbf{0.87} & \textbf{0.98} & \textbf{0.80} & \textbf{0.89} & 0.64\\
		& CGC & -- & 0.72 & 0.93 & 0.43 & 0.59 & 0.32\\
		& {DAGMA} & 0.61 & 0.17 & 0.33 & 0.17 & 0.22 & 0.25\\
		& MCMC & 0.127 & 1 & 0 & -- & 0 & 0.68\\
		\hline
		$6$ & GraphEM & 0.097 & 0.68 & 0.38 & 0.70 & 0.49 & \textbf{0.21}\\
		& SpaRJ & \textbf{0.078} & \textbf{0.75} & 0.53 & \textbf{0.81} & 0.63 & 3.2\\
		& CGC & -- & 0.63 & 0.65 & 0.32 & 0.45 & 1.7\\
		& {DAGMA} & 0.26 & 0.25 & \textbf{0.96} & 0.71 & \textbf{0.82} & 0.62\\
		& MCMC & 0.152 & 1 & 0 & -- & 0 & 3.5\\
		\hline
		$12$ & GraphEM & 0.102 & \textbf{0.76} & 0.34 & 0.88 & 0.49 & \textbf{0.92}\\
		& SpaRJ & \textbf{0.074} & 0.60 & 0.53 & \textbf{0.88} & 0.65 & 14.7\\
		& CGC & -- & 0.75 & 0.57 & 0.43 & 0.49 & 6.5\\
		& {DAGMA} & 0.16 & 0.25 & \textbf{0.97} & 0.86 & \textbf{0.91} & 1.0\\
		& MCMC & 0.124 & 1 & 0 & -- & 0 & 15.0\\
		\hline
	\end{tabular}
	
	\label{tab:systemtypesnisounknown}
\end{table}

We see in Table~\ref{tab:systemtypesnisounknown} that our method performs well under these challenging conditions, consistently outperforming existing methods. 
Note that the CGC and {DAGMA} metrics are unchanged from the previous section, as these methods does not require accept estimate for $\Q$.
{The deterioration of metrics is expected in this experiment, as we are inferring both the value of $\Q$ and the value of $\A$ from the same data, with the estimation of $\A$ being conditional on the estimated value of $\Q$. 
However, the sparsity structures of the estimates are better than comparable methods, and the parameter value is still well estimated, as evidenced by the RMSE value.}

{}

\subsubsection{Variable levels of sparsity}
\label{sec:numvarsparse}
{We now explore the performance of our method under variable levels of sparsity.}
{To facilitate direct comparison, all other parameters of the state-space model remain the same between sparsity levels, as well as the matrix from which $\A$ is generated. This experiment is performed on a $4\times4$ transition matrix, with $\Q = \R = \Id_4$, $\P = 10^{-8}\Id_{4}$, $T = 50$, and $\lambda = 0.5$.}

\begin{figure}[H]
	\centering
	\includegraphics[width=0.45\textwidth]{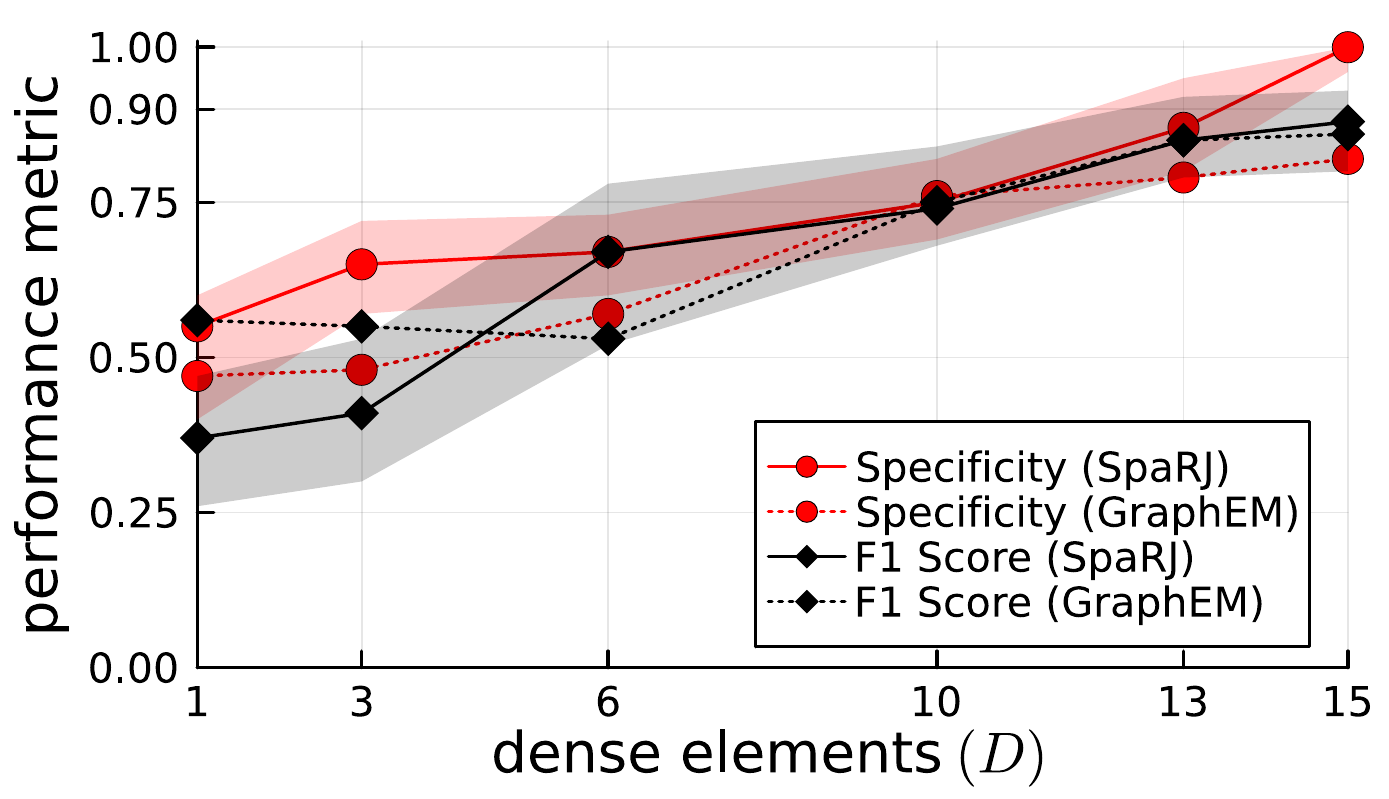}
	\caption{Sparsity metrics over variable sparsity in a $4 \times 4$ system. Shaded regions denote $95\%$ HPDIs, markers denote means. The dotted line indicates the mean performance of GraphEM.}
	\label{fig:varspar}
\end{figure}

Note that, in systems with many dense elements, the effect of each element can be emulated by changing the values of a number of other elements, making sparsity recovery difficult in these cases. 
Our algorithm performs well in general, consistently outperforming other methods in this case.

{We observe from Figure~\ref{fig:varspar} that both methods generally perform better as the number of sparse elements increases.
This is due to sharper likelihood changes occurring when sparsity changes when assessing these models. 
For particularly dense transition matrices, GraphEM outperforms SpaRJ, due to the model proposal step of SpaRJ requiring more transitions to walk the larger space.
This could be remedied with a prior encoding more information for these sampling regimes.
For example, a penalty relating to the number of sparse elements could be incorporated into the prior.
This ease of encoding prior preference in the model space is a strength of SpaRJ, and is not possible in comparable methods.
{Furthermore, SpaRJ can be assisted by GraphEM via the provision of a sparse initial value $\A_0$, which will greatly speed convergence.
We have not done this for any of the numerical experiments, but in practice we recommended doing so.}}

\subsection{Application to global temperature data}
We now apply our method to real data. 
We use the average daily temperature of 324 cities from 1995 to 2021, curated by the United States Environmental Protection Agency \cite{udaytd}. 
We subset {the data} to the cities of London (GB), Paris (FR), Rome (IT), Melbourne (AU), Houston (US), and Rio de Janeiro (BR) in 2010. 
We subset to a single year to avoid missing data. 

\begin{figure}[ht]
    \subcaptionbox{Graph recovered via GraphEM.\label{fig:realdatagem}}
         {
         \centering
         \fbox{\includegraphics[width=0.215\textwidth]{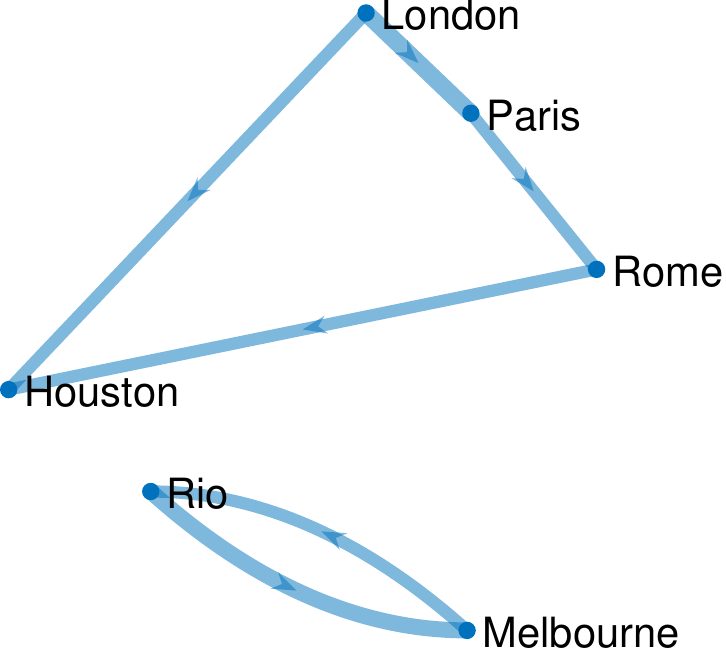}}
         }
         \hspace{-1em}
     \subcaptionbox{Graph recovered via SpaRJ.\label{fig:realdata}}
         {
         \centering
         \fbox{\includegraphics[width=0.215\textwidth]{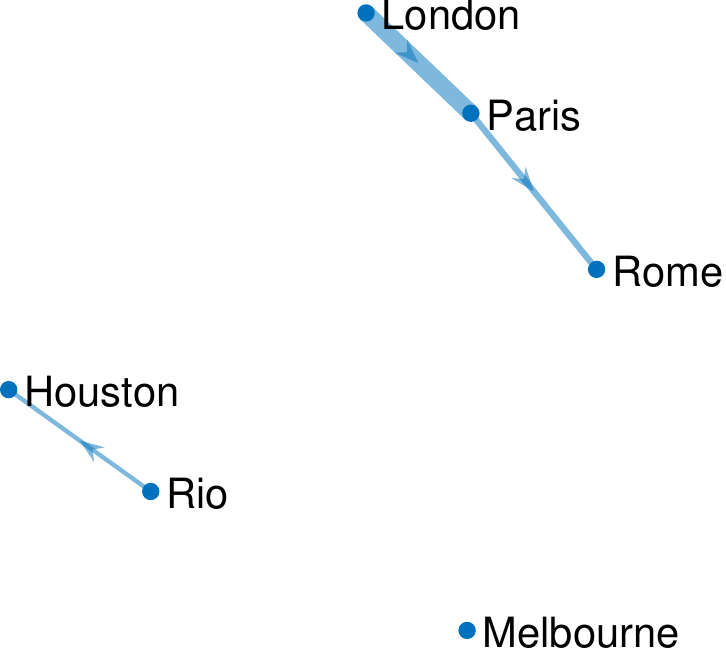}}
         }
	\caption{Graphs of state relationships recovered from temperature data using both GraphEM and SpaRJ. Self-self edges are omitted for clarity, but present in all instances {(i.e., the elements in the diagonal of $\A$ are different from zero)}. }
\label{fig:rdboth}
\end{figure}

We set the parameters as follows: $\pi_{-1} = 0.5, \pi_{0} = 0.8, \lambda = 0.5,$ and $\lambda_j = 0.2$.
We estimate $\Q$ using the EM scheme detailed in Section~\ref{sec:numunknown}, and set $\H = \Id_6, \R = 0.5\Id_6$ as per the data specification. 
{The results for GraphEM are given graphically in Figure~\ref{fig:realdatagem}, while Figure~\ref{fig:realdata} displays the results for SpaRJ. 
In Figure~\ref{fig:rdboth}, edge thickness for GraphEM is proportional to the number of times an edge appears in $100$ independent runs, whereas for SpaRJ edge thickness is proportional to the number of post-burnin samples from $100$ chains with the edge present.}

{It is well known that weather phenomena are highly non-linear, and are driven by both local and global factors.
Not all driving factors are recorded in the data, therefore making it a challenging task to extract statistically causal relationships between locations.
For example, neither barometric pressure nor rainfall are utilised, which would assist with localisation \cite{bauer2015quiet}.} 

We see that the geographical relationships between the cities are well recovered by SpaRJ. 
We see that the European cities form one cluster, Melbourne is separate, and Rio weakly affects Houston in Figure~\ref{fig:realdata}. 

{Second, note that there exists no ground truth to compare against in this problem. 
GraphEM recovers the graph given in Figure~\ref{fig:realdatagem}, which does not reflect the geographical positioning of the cities, as there are connections across large distances which is not physically reasonable. 
In addition, this graph cannot be interpreted probabilistically, as is possible with SpaRJ. On the other hand, GraphEM is generally faster compared to SpaRJ.}

{The SpaRJ estimate, presented in Figure~\ref{fig:realdata}, offers the capacity for additional inference. 
For instance, in Figure~\ref{fig:realdatagem}, all edges recovered by GraphEM are of a similar thickness, indicating that they are recovered by many independent runs of the algorithm.
This is a desirable characteristic of GraphEM, as it is indicative of good convergence, although it does not admit a probabilistic interpretation of state connectivity.}

{SpaRJ recovers the edges probabilistically, which is made apparent in Figure~\ref{fig:realdata} by the variable edge thicknesses. 
In SpaRJ, the number of post-burnin samples in which a given edge is present gives an estimate of the probability that this edge is present.
This is of particular interest when inferring potential causal relationships, as in this example.
This property follows from the broader capacity of SpaRJ to provide Monte Carlo uncertainty quantification. 
For example, a credible interval for the probability of an edge being present can easily be obtained via bootstrapping with the output of SpaRJ, which is not possible with GraphEM, as it is designed to converge to a point.}

{We used a value of $\lambda = 1.2$ in GraphEM for this estimation. However, increasing $\lambda$ would make the self-self edges disappear (i.e., zeros would appear in the diagonal of $\A$ before edges among cities would be removed).}
{Comparing the results in Figure~\ref{fig:rdboth}, we see that the output from SpaRJ is more feasible when accounting for geophysical properties. 
It is not reasonable for the weather of cities to affect each other across very large distances and oceans over a daily timescale, and the parameter estimate should reflect this. 
This spatial isolation is present in the SpaRJ estimate in Figure~\ref{fig:realdata}, but is absent from the GraphEM result in Figure~\ref{fig:realdatagem}.
}

\subsection{Assessing convergence}

{As our method is a MCMC method, it is not desirable to converge in a point-wise sense, however it is desirable to converge in distribution to the target distribution.
However, we cannot use standard metrics such as $\hat{R}$ \cite{gelman2013bayesian} to assess convergence, as these assume that the same parameters are being estimated at all times, which is not the case in our method.}
{In the literature there are specific methods to assess convergence for RJMCMC algorithms, with proposed methods \cite{brooks1998convergence} breaking down for large model spaces with few visited models (as is the case here).
However, due to the tight linking between the model space and the parameter space, we can assess convergence via a combination of model metrics and parameter statistics \cite{robert2013monte}.}

{In order to properly assess convergence, we must take into account both the model space and parameter space, and assess convergence in both.
As our model space is closely linked to the values in the parameter space, we are able to assess convergence by jointly observing parameter and model metrics.}

\begin{figure}[ht] 
	\centering
	\includegraphics[width=0.48\textwidth]{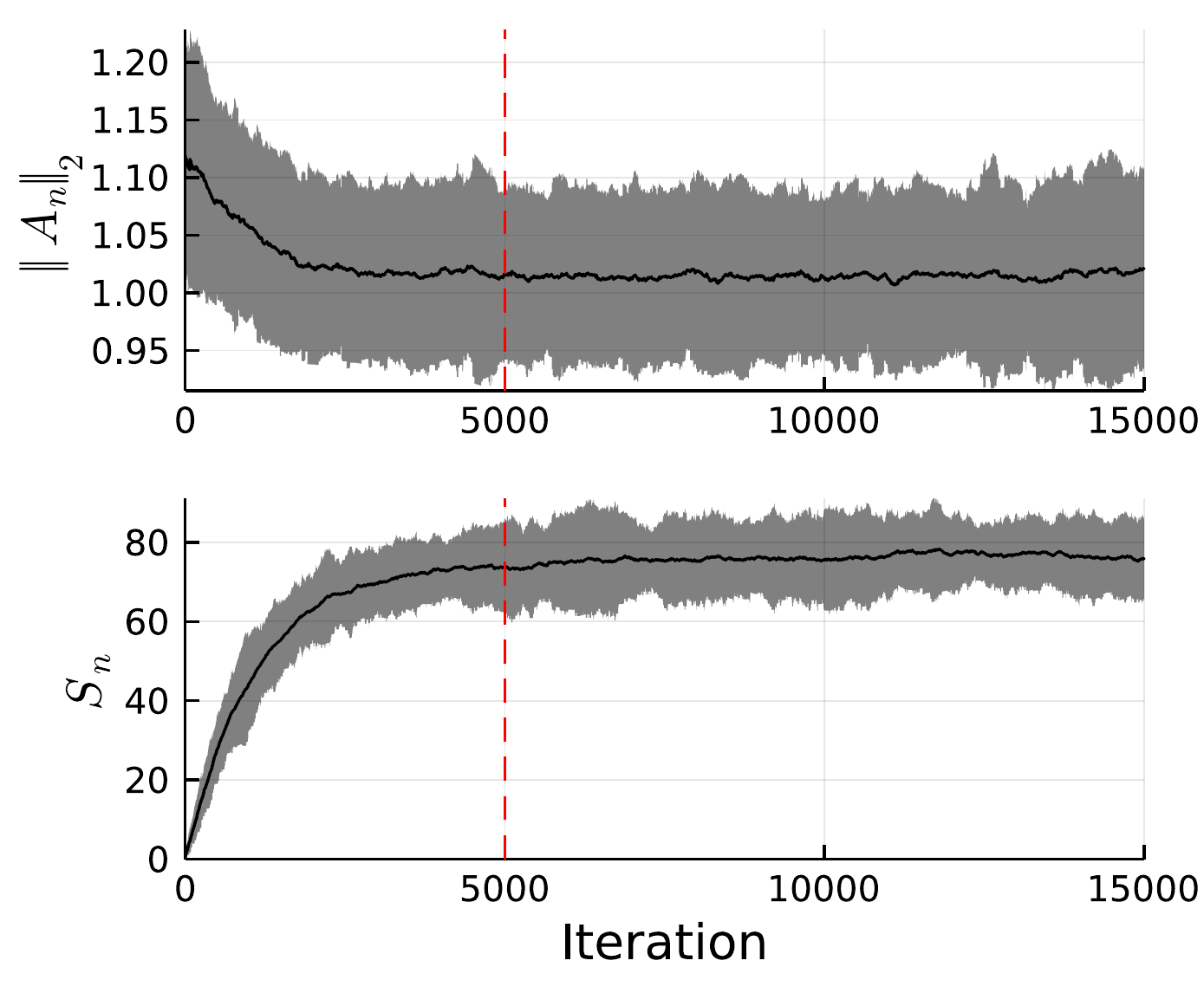}
	\caption{Progression of sample metrics in the $12 \times 12$ estimated state covariance model. The red line indicates the end of the burn-in period. Shaded regions are $95\%$ HPDIs. Black line indicates the mean.}
	\label{fig:convres}
\end{figure}

We track both the spectral norm of the sampled $\A_n$ (parameter metric) and the number of sparse elements (model metric), {and plot them in Figure~\ref{fig:convres}.}
We observe that convergence in both the parameter space and the model space occur quickly, and that convergence seems to occur before the burn-in period ends. 
This is the case for all examples, with the {exemplar system} being the slowest to converge. 
It is possible to decrease the time to convergence in several ways, such as better estimates of the model parameters.
However, parameters such as $\pi_0$ and $\pi_{-1}$ will also alter the speed of convergence, although the manner in which they do so is dependent on the true dynamics.
{We note that convergence speed is faster for lower dimensional $\A$, and conversely is slower for larger $\A$. 
This is due to a larger $\A$ matrix having more variables to estimate, and hence requiring sampling from a higher dimensional space. 
Furthermore, the sampler converges faster for longer series lengths.}
Experimentally, sparser models benefit from a larger $\pi_{-1}$, whereas denser models benefit from a smaller $\pi_{-1}$.
Increasing $\pi_0$ increases convergence speed in the parameter space, but decreases convergence speed in the model space.
Convergence could also be improved by using a gradient-based sampler for the parameter posterior, as the gradient is available in closed form \cite{sarkka2013bayesian}.
We find that the increase in computational cost and the reduction in modularity is not worth the increased speed of convergence.
{Finally, note that our method inherits the convergence guarantees of RWMH and RJMCMC, and therefore for a finite dimensional parameter space we are guaranteed to converge to the target sampling distribution given sufficient iterations.}
\section{Conclusion}
\label{sec:conclusion}
{In this work we have proposed the SpaRJ algorithm, a novel Bayesian method for recovering sparse estimates of the transition matrix of a linear-Gaussian state-space model.
In addition, SpaRJ provides Bayesian uncertainty quantification of Granger causality between state elements, following from the interpretation of the transition matrix as representing information flow within an LGSSM.
The method, built on reversible jump Markov chain Monte Carlo, has strong theoretical guarantees, displays performance exceeding state-of-the-art methods in both challenging synthetic experiments and when operating on real-world data, and exhibits great potential for extension.} 

\appendix
\subsection{{Guidance for choice of parameters}}
\label{app:paramhints}
\begin{table}[ht]
	\scriptsize
    \caption{{Hints for choosing parameter values}}
    \label{tab:phints}
	\begin{center}
		\begin{tabular}{|c|c||c|}
		    \hline
		    Parameter & Hint & Recommended value(s)\\
			\hline
			\hline
			$\lambda$ & Increase to promote sparsity & $\in[\exp(-2), \exp(2)]$\\
			\hline
			$\lambda_j$ & Decrease to increase acceptance rate & $\in [0, 0.5], \approx 0.2$, \\
			\hline
			$\pi_0$ & Increase to increase acceptance rate & $\in[0.75, 0.99], \approx 0.8$\\
			\hline
			$\pi_{-1}$ & Increase to promote sparsity & $\in[0.4,0.6], \approx 0.5$\\
			\hline
			$\sigma, \sigma_c$ & Decrease to increase acceptance rate & $\in[0.05, 0.15], \approx0.1$\\
			\hline
		\end{tabular}	\end{center}
\end{table}
\subsection{Truncated Poisson distribution}
\label{app:TPoi}
Denote the Poisson distribution with rate $\lambda$ that is left-truncated at $a$ and right-truncated at $b$ by $\text{TPoi}(\lambda, a, b)$. This distribution has support $n \in \mathbb{N} \cap [a,b]$, and probability mass function of
\begin{equation}
	\mathrm{TPoi}(n; \lambda, a, b) = \frac{\lambda^ne^{-\lambda}}{Z\cdot n!}, \text{ with } Z = \sum_{n=a}^{b}\frac{\lambda^ne^{-\lambda}}{n!}.
\end{equation}
\subsection{Correction terms}
\label{app:corr}
In order to maintain detailed balance in the sampling chain, we must account for the unequal model transition probabilities, which is done via a correction term. 
These terms arise from the RJMCMC acceptance probability, 
\begin{equation}
    \frac{l^\prime}{l} \frac{\pi_{n+1,n}}{\pi_{n,n+1}}\frac{g(u_{n})}{g(u_{n+1})}\left|\frac{\partial T_{n,n+1}(\btheta_n, u_{n})}{\partial(\btheta_n, u_{n})}\right|,
\end{equation}
in which ${\pi_{n+1,n}}/{\pi_{n,n+1}}$ is the ratio of the probability of the reverse jump to that of the forward jump. 
All other terms in the acceptance ratio are calculated in Algorithm~\ref{alg:gensparthreestep}, with the Jacobian term ignored as per Section~\ref{sec:bijec}.

As we are using log likelihoods and log acceptance ratios, we compute our correction on the log scale.
For a given jump distance $J$, we denote this log correction term $c_{j,J}$, with $j=s$ when jumping sparser, and $j=d$ when jumping denser. 
This term is equal to the $\log({\pi_{n+1,n}}/{\pi_{n,n+1}})$ term in the acceptance ratio in Step 3 of Algorithm~\ref{alg:gensparthreestep}.
The calculations for both sparser jumps and denser jumps proceed similarly, thus we detail only the derivation for sparser jumps.

The forward jump is a jump sparser, which occurs with probability $\pi_{-1}$.
When jumping sparser we truncate the jump distribution at $D_n$.
Hence, the probability of drawing a given jump length $J$ for the jump distance is $\mathrm{TPoi}(J; \lambda, 1, D_n)$.
Given $J$, the probability of choosing a given set of elements in the forward jump is $\binom{D_n}{J}^{-1}$.
Multiplying these terms we obtain 
\[\pi_{n,n+1} = \pi_{-1} \ \mathrm{TPoi}(J; \lambda, 1, D_n) \ J! \ (D_n-J)!(D_n!)^{-1}.\]

The reverse jump is a jump denser, which occurs with probability $1-\pi_{-1}$.
When jumping sparser, we truncate the jump distribution at $S_n+J$, the number of sparse elements after the forward jump occurs.
Hence the probability of drawing a given $J$ for the jump distance is $\mathrm{TPoi}(J; \lambda, 1, S_n + J)$.
Given $J$, the probability of choosing a given set of elements in the reverse jump is $\binom{S_n+J}{J}^{-1}$.
Multiplying these terms we obtain 
\[\pi_{n+1,n} = {(1-\pi_{-1}) \mathrm{TPoi}(J; \lambda, 1, S_n+J) J! S_n!}(S_n+J)!^{-1}.\]

From which we obtain the acceptance ratio
\begin{equation*}
	\frac{\pi_{n+1,n}}{\pi_{n,n+1}} = \frac{(1-\pi_{-1})\mathrm{TPoi}(J; \lambda, 1, S_n+J)S_n!\,D_n!}{\pi_{-1}\mathrm{TPoi}(J; \lambda, 1, D_n)(S_n+J)!\,(D_n-J)!},
\end{equation*} 
Writing $r := (1-\pi_{-1})(\pi_{-1})^{-1}$ we then have
\begin{equation}
	\exp(c_{\text{s},J}) = r\frac{\mathrm{TPoi}(J; \lambda, 1, S_n+J)S_n!\,D_n!}{\mathrm{TPoi}(J; \lambda, 1, D_n)(S_n+J)!\,(D_n-J)!},
\end{equation} 
with the corresponding term for the denser jump being 
\begin{equation}
	\exp(c_{\text{d},J}) = \frac{1}{r}\frac{\mathrm{TPoi}(J; \lambda, 1, D_n+J)S_n!\,D_n!}{\mathrm{TPoi}(J; \lambda, 1, S_n)(D_n+J)!\,(S_n-J)!}.
\end{equation}

In the case of jumping to and from maximal density (MD) and maximal sparsity (MS) further adjustment is required. 
In these cases we replace $r$ following Table~\ref{tab:rmod}.
\begin{table}[ht]
	\footnotesize
    \caption{$r$ terms for edge cases.}
    \label{tab:rmod}
	\begin{center}
		\begin{tabular}{|c||c|c|c|c|}
			\hline
			Jump & to MS & to MD & from MS & from MS \\
			\hline
			$r$ & $(\pi_{-1})^{-1}$ & $1-\pi_{-1}$ & $(\pi_{-1})^{-1}$ & $1-\pi_{-1}$ \\
			\hline
		\end{tabular}
	\end{center}
\end{table}

\vspace{-2em}
\balance
\bibliographystyle{IEEEtran}


\end{document}